\documentclass[journal]{new-aiaa} 
\usepackage[utf8]{inputenc}
\usepackage{multirow}
\usepackage{graphicx}
\usepackage{tablefootnote}
\usepackage{subcaption}
\usepackage{booktabs} 
\usepackage{cleveref}
\usepackage{amsmath}
\usepackage[version=4]{mhchem}
\usepackage{siunitx}
\usepackage{amsfonts}       
\usepackage{nicefrac}       
\usepackage{microtype}      
\usepackage{xcolor}         
\usepackage{algorithm}
\usepackage{algpseudocode}
\usepackage{framed}
\usepackage{MnSymbol}
\usepackage{comment}
\usepackage{xargs}                      
\usepackage{pgf, tikz}
\usepackage{pgfplots}
\pgfplotsset{compat=1.18}
\usepackage{diagbox}
\usetikzlibrary{shapes.geometric, arrows.meta, positioning}

\tikzstyle{startstop} = [rectangle, rounded corners, minimum width=2.8cm, minimum height=1.0cm, text centered, draw=black, fill=gray!10]
\tikzstyle{process} = [rectangle, minimum width=4.0cm, minimum height=1.0cm, text centered, text width=4.5cm, align=center, draw=black, fill=blue!5]
\tikzstyle{decision} = [diamond, aspect=3, text width=3.2cm, align=center, text centered, draw=black, fill=orange!10]
\tikzstyle{arrow} = [thick,->,>=stealth]
\newcommand{\changed}[1]{\textcolor{black}{#1}}

\usepackage{longtable,tabularx}
\title{Autonomous Air-Ground Vehicle Operations Optimization in Hazardous Environments: A Multi-Armed Bandit Approach}

\author{Jimin Choi\footnote{Ph.D. Candidate, Department of Aerospace Engineering\changed{; jiminch@umich.edu. AIAA Student Member (Corresponding Author).}} and Max Z. Li\footnote{Assistant Professor, Department of Aerospace Engineering, Department of Civil and Environmental Engineering, Department of Industrial and Operations Engineering, University of Michigan, Ann Arbor, Michigan 48109, AIAA Member.}}
\affil{University of Michigan, Ann Arbor, MI 48109, USA}

\begin{document}

\maketitle

\begin{abstract}
Hazardous environments such as chemical spills, radiological zones, and bio-contaminated sites pose significant threats to human safety and public infrastructure. Rapid and reliable hazard mitigation in these settings often unsafe for humans, calling for autonomous systems that can adaptively sense and respond to evolving risks. This paper presents a decision-making framework for autonomous vehicle dispatch in hazardous environments with uncertain and evolving risk levels. The system integrates a Bayesian Upper Confidence Bound (BUCB) sensing strategy with task-specific vehicle routing problems with profits (VRPP), enabling adaptive coordination of unmanned aerial vehicles (UAVs) for hazard sensing and unmanned ground vehicles (UGVs) for cleaning. Using VRPP allows selective site visits under resource constraints by assigning each site a visit value that reflects sensing or cleaning priorities. Site-level hazard beliefs are maintained through a time-weighted Bayesian update. BUCB scores guide UAV routing to balance exploration and exploitation under uncertainty, while UGV routes are optimized to maximize expected hazard reduction under resource constraints. Simulation results demonstrate that our framework reduces the number of dispatch cycles to resolve hazards by around 30\% on average compared to \changed{uninformed} baseline dispatch strategies, underscoring the value of uncertainty-aware vehicle dispatch for reliable hazard mitigation.
\end{abstract}

\section{Nomenclature}
\setcounter{table}{-1}
{\renewcommand\arraystretch{1.0}
\noindent\begin{longtable*}{@{}l @{\quad=\quad} l@{}}
$ i, j $        & Index for candidate sites \\
$ t $           & Index for discrete time step or round \\
$ k $           & Index for observations at a site \\
$ m $           & Index for vehicle (UAV or UGV) \\
$ S $           & Set of candidate sites \\
$ M $           & Set of vehicles \\
$ G = (V, E) $  & Graph of depot and site locations, with edges $E$ \\
$ \rho_i $      & Intrinsic hazard growth rate at site $i$ \\
\changed{$ \rho_{\max}$}   & \changed{Upper bound on intrinsic hazard growth rate}\\
$ \phi $        & Spatial influence coefficient \\
$ K $           & Hazard saturation threshold (max hazard level) \\
$ \lambda $     & Time-decay rate in Bayesian belief update \\
$ \alpha $      & Gradient smoothing factor for trend extrapolation \\
$ \beta $       & BUCB exploration–exploitation trade-off parameter \\
$ \gamma $      & Variance inflation rate for unobserved sites \\
$ \zeta $       & Uncertainty boost for residual hazard correction \\
$ \kappa $      & Distance penalty coefficient in sensing score \\
$ D_m $         & Maximum travel distance for UAV $m$ \\
$ Q_m $         & Cleaning capacity of UGV $m$ \\
$ Q_{\text{unit}} $ & Max hazard cleaned per visit at a site \\
$ \sigma_\varepsilon^2 $ & Variance of sensor measurement noise \\
$ \sigma_{\max}^2 $ & Saturation limit for uncertainty \\
$ H_{i,t} $     & True hazard level at site $i$ at time $t$ \\
$ \mu_{i,t} $   & Estimated hazard mean at site $i$ at time $t$ \\
$ \sigma^2_{i,t} $ & Estimated variance of hazard at site $i$ at time $t$ \\
$ y_{i,k} $     & Noisy observation of hazard at site $i$ \\
$ \tau_{i,k} $  & Time of observation $k$ at site $i$ \\
$ g_{i,t} $     & Estimated hazard trend at site $i$ \\
$ x_{im} $      & Binary variable: 1 if site $i$ is visited by vehicle $m$ \\
$ z_{ijm} $     & Binary variable: 1 if vehicle $m$ travels from site $i$ to $j$ \\
$ u_{im} $      & Subtour elimination variable (MTZ constraint) \\
$ S_{i,t} $     & Raw BUCB score at site $i$ \\
$ \tilde{S}_{i,t} $ & Distance-adjusted BUCB score at site $i$ \\
$ \pi_i^{\text{sense}} $ & Visit value for sensing VRPP \\
$ \pi_i^{\text{clean}} $ & Visit value for cleaning VRPP \\
$ R_{i,t} $     & Expected hazard removed if site $i$ is cleaned \\
$ \delta_i $    & Cleaning demand at site $i$ \\
$ T_{\text{end}} $ & Termination round (when all hazards are mitigated) \\
$ \mathcal{J} $  & Cumulative hazard objective over time \\
\end{longtable*}

\section{Introduction}

In recent years, the deployment of autonomous vehicles for site monitoring and maintenance has gained significant attention due to their ability to cover large areas and efficiently access hard-to-reach locations. Especially, unmanned aerial vehicles (UAVs) offer flexibility and autonomy that traditional modalities cannot match, making them ideal for environmental monitoring, infrastructure inspection, agricultural surveillance, and disaster management applications \cite{shakhatreh2019unmanned}. Among various UAV dispatch applications, this paper focuses on scenarios where UAVs detect and clean contaminants in hazardous environments, such as chemical spill sites or areas with radioactive contamination. According to 2024 statistics by HAZWOPER-OSHA, workers in hazardous-waste operations face about 5.7 workplace injuries per 1,000 workers annually, driven largely by falls, chemical exposures, and heavy equipment accidents \cite{hazwoper_osha_2025}. These persistent risks underscore the need for approaches that minimize direct human exposure in such environments. Recognizing this, the potential of using machines rather than humans to eliminate hazards has been explored since the 1990s \cite{jaselskis_robotic_1994}. Recent advancements in UAV technology have proven their suitability for addressing hazards \cite{kas_using_2020}. However, one key challenge in such environments is that these hazards can fluctuate over time due to various factors such as environmental conditions or inherent hazard characteristics. The uncertainty and variability of these hazards necessitate real-time and adaptive strategies for resource allocation, which traditional static methods often fail to address.

Recent field deployments further underscore the value of autonomous systems in hazardous scenarios. In 2020, DJI enterprise UAVs such as the Matrice 300 RTK were deployed during catastrophic floods and landslides in Vietnam to assist rescue teams with thermal imaging, aerial reconnaissance, and 3D terrain mapping in inaccessible areas \cite{dji_vietnam_2021}. Similarly, in offshore oil spill response exercises, Terra Drone Angola utilized UAVs equipped with thermal and optical sensors to detect and map contaminated regions rapidly, enabling faster and safer decision-making during environmental emergencies \cite{terra_drone_angola_2019}. On the ground, Rheinmetall’s Mission Master family of unmanned ground vehicles has demonstrated multi-role capabilities such as logistics support, medical evacuation, and reconnaissance under extreme terrain and weather conditions \cite{rheinmetall_mission_master}. These examples illustrate the growing feasibility of deploying integrated autonomous systems for coordinated sensing and mitigation in dynamic, high-risk environments.

To operate effectively in such settings, autonomous systems must reason under partial observability and dynamically evolving risks since hazardous environments are complex and not static. Risk levels may increase due to weather effects, diffusion, or latent failures, and sensor measurements are often noisy and incomplete \cite{8877114}. In this context, static response plans based on pre-defined priorities or fixed routes can result in delayed detection, suboptimal intervention, or inefficient resource use. These challenges necessitate decision-making frameworks that can dynamically balance the competing needs of hazard detection and mitigation. Specifically, systems are required to resolve the exploration–exploitation trade-off by deciding whether to allocate limited sensing resources to under-observed sites to reduce uncertainty (exploration), or to focus on known high-risk locations for immediate intervention (exploitation). This trade-off is especially difficult when vehicles face routing and capacity constraints, and when hazard levels evolve continuously over time. 

We address these challenges by proposing a decision-making framework that integrates uncertainty-aware sensing with task-constrained vehicle routing \changed{\cite{choi2025bandit}.} The method coordinates UAV-based sensing and unmanned ground vehicle (UGV)-based cleaning through a principled combination of Bayesian Upper Confidence Bound (BUCB) scoring and vehicle routing problems with profits (VRPP). By modeling evolving hazard beliefs and optimizing the vehicle routes, the system prioritizes sensing and cleaning actions adaptively to maximize overall hazard reduction efficiency. \changed{We evaluate the proposed framework against baseline sensing strategies, including Random and Round-Robin approaches, as well as a Point-Estimate variant that removes uncertainty-aware exploration from BUCB.}

\section{Literature Review}
\subsection{UAV Applications in Hazardous Environments}
The application of UAVs in hazardous environments has been explored recently due to their potential to reduce human exposure to dangerous conditions. However, extensive research in this area has not yet been conducted, leaving significant room for further investigation and development. We first review recent works on hazardous waste collection from the perspective of UAV routing problems: Kaabi et al. present a two-phase approach for hazardous waste collection using UAVs, emphasizing planning and operational strategies to address constraints like battery endurance and load capacities \cite{kaabi_2-phase_2023}. Abdulsattar et al. expanded on this by integrating a custom heuristic with Ant Colony Optimization (ACO) \cite{abdulsattar_ant_2023}. The study demonstrates that this hybrid outperforms previous approaches in minimizing collection time across varying problem sizes. Harrath et al. develop a priority rule-based algorithm to optimize multi-UAV routing for hazardous waste collection, where UAVs are assigned based on shortest, longest, or median flying times \cite{harrath_algorithm_2024}. These works illustrate the emerging potential of UAV-based solutions for hazardous environments, while also highlighting the need for more comprehensive, adaptive planning frameworks.

\subsection{Coordinating UAVs and UGVs for Integrated Operations}
The coordination of UAVs and UGVs has emerged as an effective strategy for tasks that demand both the agility of aerial platforms and the endurance and payload capacity of ground assets \cite{liu2022review}. Several studies show how the two systems work together in practice. Lazna et al. used UAVs for rapid mapping and preliminary radiation measurements before delegating precise localization to UGVs \cite{doi:10.1177/1729881417750787}, and De Petrillo et al. addressed search planning for UAV–UGV teams operating in subterranean environments with significant localization uncertainty \cite{9448603}. In addition, Ren et al. introduced a mission-centered collaboration framework based on situation awareness \cite{10.1007/978-981-16-9492-9_334}. At the same time, Klodt et al. developed a visibility-constrained motion control approach to ensure UAVs maintain visual contact with UGVs while providing critical information \cite{7320804}. These studies demonstrate the strong potential of UAV–UGV coordination for complex and dynamic operations, including those in hazardous environments.

\subsection{Vehicle Routing for Waste and Hazard Collection}
In the broader context of vehicle routing problems (VRP) for cleaning tasks, significant advancements have been made in the last decade \cite{liang_waste_2022}. Kim et al. addressed the vehicle routing problem with time windows (VRPTW) for commercial waste collection, considering real-world constraints such as driver lunch breaks and multiple disposal trips \cite{kim_waste_2006}. Dotoli and Epicoco proposed an optimization technique for the vehicle routing and scheduling problem focused on hazardous waste collection and disposal \cite{dotoli_vehicle_2017}. Rabbani and his team presented a multi-objective metaheuristic algorithm to solve hazardous waste collection problems, incorporating service time windows and workload balancing \cite{rabbani_using_2021}. Suksee and Sindhuchao developed a heuristic called GRALNSP to address location and routing issues for infectious waste collection in Northeast Thailand \cite{suksee_grasp_2021}. Recent extensions of the VRP framework have explored mixed-vehicle systems, such as the Truck-Drone Routing Problem (TDRP) and Flying Sidekick traveling salesman problem (TSP), primarily in last-mile delivery contexts \cite{liang2022survey, DEFREITAS201895}. While these models were not originally designed for hazardous environments, they offer a valuable basis for integrating aerial and ground platforms in coordinated routing strategies. Although these extensions offer promising coordination strategies, they do not address the balance between exploration and exploitation, which is important in dynamic hazardous environments. This trade-off is critical in uncertain environments where the information about site conditions evolves. While a recent study investigates bi-level UAV routing based on expected information gain \cite{bilevel2025}, its scope is limited to monitoring,  leaving integrated planning for hazard mitigation unaddressed. Ultimately, addressing this trade-off requires decision-making models that integrate real-time learning with routing under uncertainty, going beyond traditional VRP formulations.

\subsection{Multi-Armed Bandits under Uncertainty}
To address this, our research utilizes a multi-armed bandit (MAB) approach that effectively balances exploration and exploitation in uncertain environments \cite{MAL-068}, within the broader context of uncertainty-aware sequential decision-making \cite{choi_adaptive_rmdp_2026}. Le Ny et al. demonstrated the use of restless bandit allocation indices for multi-UAV dynamic routing under partial observations, highlighting the efficacy of MAB algorithms for operating with incomplete information \cite{le_ny_multi-uav_2008}. Lagos and Pereira extended the application of MAB through hyper-heuristics for combinatorial optimization problems; their study highlighted how adaptive algorithms could dynamically choose the most effective heuristics based on real-time performance feedback, demonstrating robustness across a variety of optimization contexts \cite{lagos_multi-armed_2024}. Recently, MAB approaches have been applied in disaster settings for dynamic resource planning \cite{smartcities8010005}, UAV-assisted communication \cite{8669870}, and energy-aware trajectory optimization \cite{s23031402}, demonstrating their effectiveness in adaptive planning under uncertainty. \changed{Building on these ideas, our framework integrates uncertainty-aware bandit-based prioritization with VRPP to support coordinated hazard sensing and cleaning under partial observability.}

\section{Contributions of Work} 
Our research contributions are as follows: 

\begin{enumerate}
    \item We propose a decision-making framework that integrates a BUCB strategy \changed{\cite{pmlr-v22-kaufmann12}} with VRPP. This integration enables dynamic allocation and routing of UAVs and unmanned ground vehicles (UGVs) in hazardous environments, allowing the system to jointly reason about exploration–exploitation trade-offs and operational constraints during both sensing and cleaning phases.
    
    \item We develop an adaptive sensing mechanism that uses time-weighted Bayesian update and smoothed gradient propagation to maintain predictive hazard beliefs over time. This approach prioritizes uncertain or rapidly changing regions and improves the reliability of sensing decisions under partial observability.

    \item We formulate task-specific VRPPs for sensing and cleaning. Sensing uses BUCB scores adjusted for distance, while cleaning uses a belief-based estimate of hazard impact. Each routing problem is solved per round under vehicle constraints, enabling route-level planning that accounts for both task priorities and operational efficiency.
                
    \item We validate the proposed framework through simulations in dynamic and stochastic environments. The results show that our method consistently outperforms baseline strategies in terms of termination rounds, completing tasks in fewer decision cycles while maintaining reliable performance under partial and noisy observations.

\end{enumerate}

\section{Problem Statement}
\label{sec:problem_statement}
\subsection{Overview and Objectives}

We consider a dynamic vehicle dispatch problem in which a set of candidate sites is dispersed across a hazardous environment. Each site represents a potential hazard source due to factors such as chemical spills, biological threats, or radioactive exposure. The true hazard level at each site is initially unknown and can only be inferred through direct, noisy observations obtained via active sensing.

A team of UAVs is deployed to monitor these sites by collecting sensor measurements, while a separate team of UGVs is tasked with performing cleaning operations to reduce hazard levels. This division of roles is illustrated in \Cref{fig:operation}. UAVs offer fast and flexible access to a wide area, making them well-suited for rapid sensing and information gathering. In contrast, UGVs are well-suited for on-site hazard cleaning as they can sustain long-duration operations and carry substantial payloads, including robotic arms for debris removal, manipulators for handling contaminated materials, and chemical reservoirs for spraying decontamination agents.

\begin{figure}[htbp]
    \centering
    \includegraphics[width=0.55\linewidth]{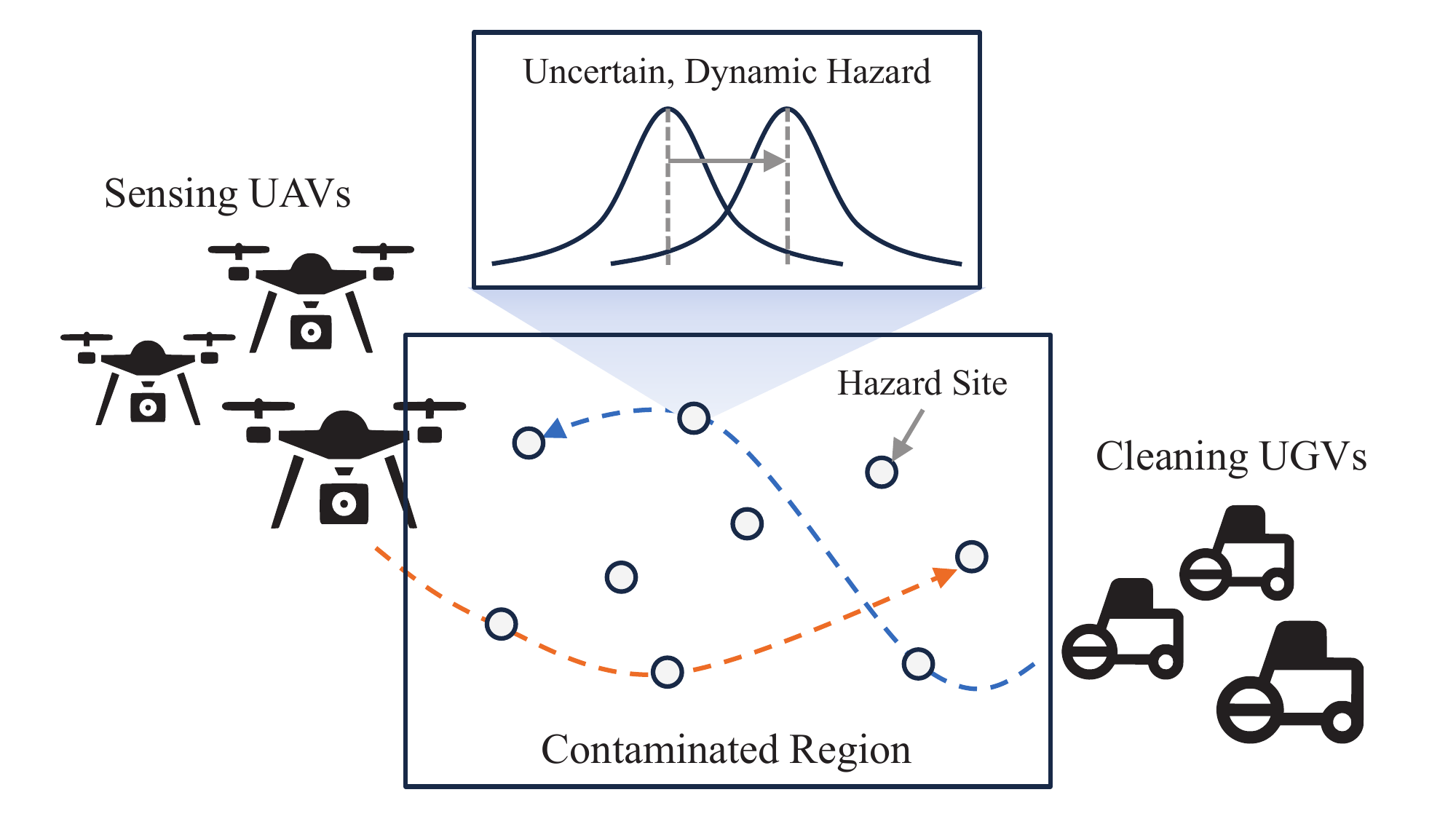}
    \caption{\changed{System schematic showing UAVs for aerial sensing and UGVs for ground-level cleaning across multiple hazardous sites with uncertain dynamics.}}
    \label{fig:operation}
\end{figure}

Each decision round consists of two sequential stages. In the sensing stage, UAVs are routed to visit a subset of sites and obtain noisy observations of current hazard levels. Based on the updated hazard estimates, UGVs are then dispatched in the cleaning stage to perform mitigation actions at selected sites. Both sensing and cleaning operations are subject to resource constraints, including a limited number of available vehicles, travel range restrictions, and bounded cleaning capacity. Each vehicle operates along a single route per round, with routes planned to maximize either information gain (for UAVs) or hazard reduction (for UGVs). 

The primary objective is to complete the task in as few rounds as possible, where each round corresponds to one full cycle of sensing and cleaning. Specifically, we aim to minimize the termination round \( T_{\text{end}} \), defined as the earliest round in which all site hazards have been fully mitigated. A formal definition is provided in \Cref{sec:evaluation_metrics}. Termination round is a suitable objective from an operational perspective, as using fewer rounds directly translates to lower resource consumption, reduced mission time, and more efficient system deployment in real-world settings. Similar to previous VRP research that frames minimizing total cost as a primary optimization goal \cite{LEE2006265, 0278cc85-5bd0-38d1-b78b-21ed2d63bafc}, our work interprets fewer termination rounds as an operational proxy for cost and resource efficiency.

\changed{While this round-based formulation abstracts away explicit travel-time and service-time constraints, it allows us to evaluate policy-level efficiency without committing to a specific temporal structure. Explicit time-criticality would naturally lead to the VRPTW formulation. Importantly, the proposed sensing–planning–cleaning framework is modular, and replacing the underlying VRPP solver with a VRPTW variant would allow mission-time constraints to be incorporated without altering the belief update or uncertainty-aware prioritization mechanisms.}

We assume that the locations and number of \( N \) candidate sites are known, but the true hazard levels \( H_{i,t} \) over time are not directly observable and must be estimated through noisy sensing observations. Since not all sites can be visited during each round, the system must prioritize its limited sensing and cleaning actions. This naturally gives rise to an exploration-exploitation dilemma: the planner must continuously decide between exploring uncertain or infrequently observed sites to improve situational awareness and exploiting current knowledge to clean locations with high estimated hazards. This challenge is further compounded by the evolving nature of hazard levels, which change over time due to intrinsic dynamics and spatial interactions.

\subsection{Sources of Uncertainty}
\subsubsection{Hazard Dynamics} \label{sec:hazard_dynamics}
Accurately modeling how such hazards evolve spatially and temporally is crucial for effective sensing and decision-making. Prior work has examined this through a variety of physical and statistical approaches, including Gaussian plume models for radioactive dispersion and CFD-based simulations for complex atmospheric releases \cite{CAO2020116925, EMST-2022-0009}. In contrast, our approach adopts a simplified spatiotemporal hazard model. This choice is motivated by the need for computational efficiency and real-time applicability in sequential decision-making settings. In our setting, a simplified model provides a tractable foundation, which could be extended with more complex dynamics in future work.

Each site within the environment possesses a latent hazard level that evolves over time. This temporal evolution is governed by both intrinsic site-specific dynamics and interactions with neighboring sites \cite{10.1007/978-3-319-12145-1_12}. Specifically, the hazard level at site \( i \) at time \( t+1 \) is modeled as,

\begin{equation}
H_{i,t+1}
= \underbrace{H_{i,t}}_{\text{previous level}}
+ \underbrace{ \rho_i \, H_{i,t} \left( 1 - \frac{H_{i,t}}{K} \right) }_{\text{logistic growth}}
+ \underbrace{ \phi \sum_{j \ne i} \frac{ H_{j,t} }{ \lVert \mathbf{p}_i - \mathbf{p}_j \rVert_2 + 1 } }_{\text{spatial influence}}
\end{equation}
where \( \rho_i \) is the intrinsic growth rate at site \( i \), \( K \) is the failure threshold representing the maximum tolerable hazard level before the site becomes inaccessible. For example, in radiological zones, this could mean radiation levels that disrupt electronics, while in chemical spill sites, it might correspond to concentrations that overwhelm protective gear. \( \phi \) controls the magnitude of spatial influence from other sites (with larger values increasing the degree of spatial coupling), and \( \mathbf{p}_i \in \mathbb{R}^2 \) denotes the spatial position of site \( i \). Distances are computed using the Euclidean ($\ell_{2}$) norm, which represents the standard straight-line distance in $\mathbb{R}^2$.

The first term captures the previous hazard level, while the second term represents intrinsic nonlinear growth following a logistic form. When the current hazard level is low, growth increases approximately linearly, modeling early-stage escalation. As \( H_{i,t} \) approaches the threshold \( K \), the growth rate slows, capturing saturation effects that may arise due to physical limitations, mitigation efforts, or environmental constraints. This formulation allows the hazard to grow quickly in the early phase while remaining bounded over time. The third term accounts for spatial interactions, where hazard levels at nearby sites diffuse or influence each other inversely proportional to their distance. This interaction captures phenomena such as spread through wind, water, or shared infrastructure. More complex models of hazard spread could also be integrated if needed. These hazard dynamics are nonlinear and spatially coupled, reflecting the complex and distributed nature of real-world contamination processes. \changed{Importantly, while this hazard growth model specifies how latent risk evolves over time, it does not determine the structure of the proposed sensing and deployment policy. Instead, hazard dynamics affect decision-making only through belief uncertainty and sensing urgency. Consequently, alternative hazard evolution models would primarily change how quickly priorities shift, while leaving the overall coordination and routing architecture unchanged.}

\subsubsection{Sensing Uncertainty}
The true hazard level at each site cannot be directly observed via sensing. Observations are corrupted by Gaussian noise, capturing the inherent limitations of real-world sensors and environmental disturbances \cite{10.1007/3-540-59496-5_337}.
\begin{equation}
    y_{i,k} = H_{i,t} + \varepsilon_{i,k}, \quad \varepsilon_{i,k} \sim \mathcal{N}(0, \sigma_\varepsilon^2),
\end{equation}
where \( y_{i,k} \) is the \( k \)\textsuperscript{th} observation at site \( i \), and \( \sigma_\varepsilon^2 \) is the known observation noise variance. Only visited sites provide new measurements, resulting in partial and noisy information about the environment.

Due to limited sensing capacity, only a subset of sites can be observed in each round. As a result, the system operates under partial observability, with most sites carrying outdated or missing information at any given time. The noisy nature of the observations further compounds this uncertainty, making accurate hazard estimation a significant challenge.

\section{Methodology}

\subsection{Proposed Framework}
\begin{figure}[htbp]
    \centering
    \includegraphics[width=0.5\linewidth]{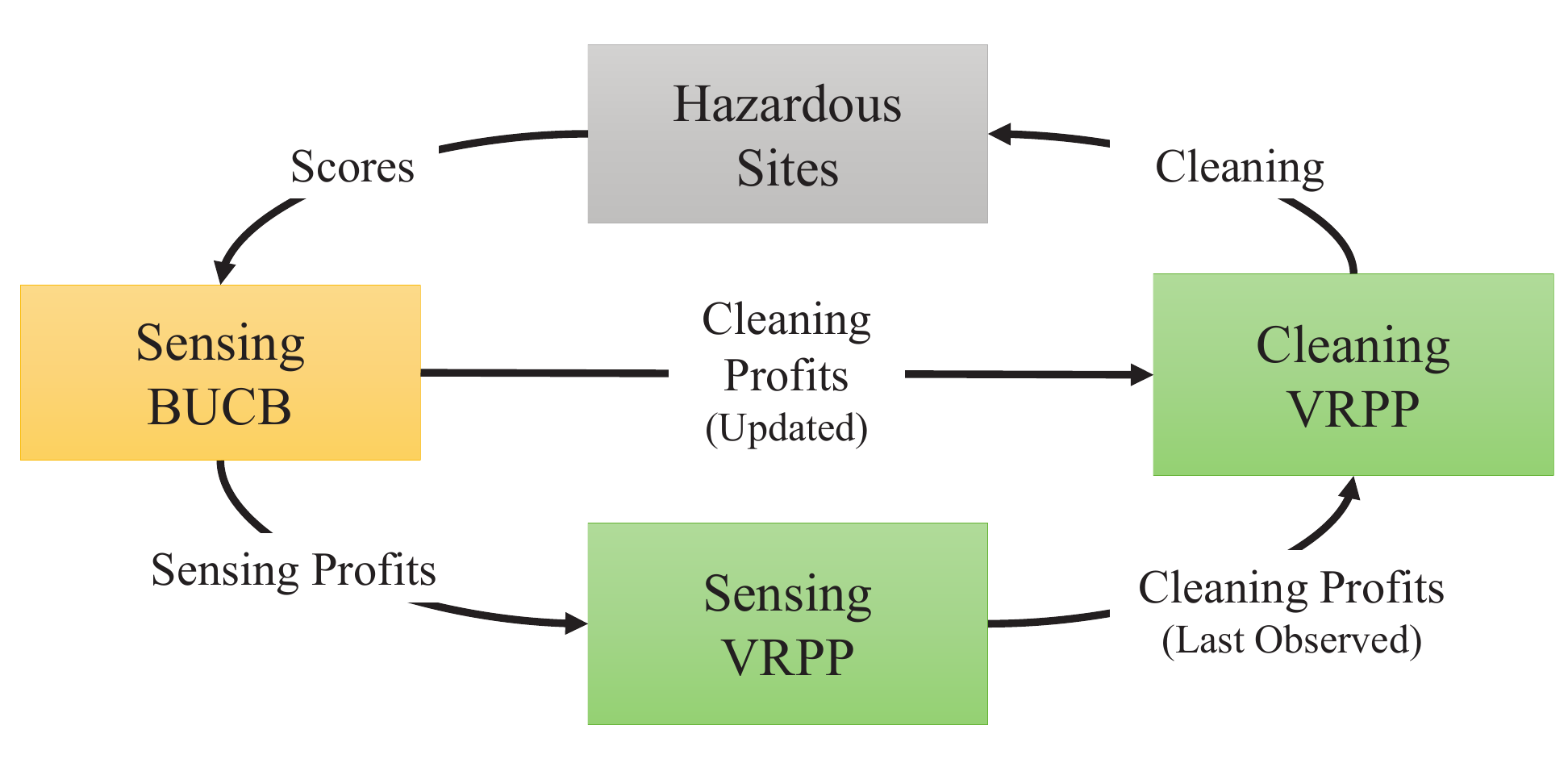}
    \caption{Vehicle dispatch workflow for hazardous environment. Sensing UAVs use BUCB to prioritize sites, while cleaning UGVs operate based on hazard estimates from the sensing.}
    \label{fig:block_diagram}
\end{figure}

\Cref{fig:block_diagram} illustrates the overall workflow of our proposed dispatch framework, which operates in repeated rounds. Each round is composed of two sequential stages: a sensing phase using UAVs and a cleaning phase using UGVs. The framework dynamically allocates limited vehicle resources to maximize long-term hazard mitigation under uncertainty.

At the start of each round, the system evaluates all candidate sites using a BUCB score. This score combines the current estimate of hazard level with the associated uncertainty, encouraging the sensing UAVs to explore sites that are either believed to be hazardous or have not been recently observed. This exploration–exploitation trade-off is central to the sensing phase, where the goal is to enhance overall situational awareness while maintaining a focus on potentially hazardous areas.

Based on these BUCB scores, the system solves a VRPP to determine efficient sensing UAV routes under distance constraints. UAVs are subsequently dispatched to a selected subset of sites to collect hazard measurements. Only the visited sites yield new information. The belief model maintains both the mean and the associated uncertainty of hazard estimates, and is updated using a time-weighted Bayesian scheme that places greater emphasis on recent observations. For sites that are not visited, uncertainty progressively increases, and hazard levels are extrapolated based on prior trends. This mechanism maintains predictive accuracy even in sparsely observed regions.

Following the sensing phase, the updated hazard estimates are used to plan cleaning routes for the UGVs. Here, we estimate the amount of hazard that could be removed from each site, given current beliefs and per-vehicle cleaning capacity limits. Another VRPP is solved to determine routes that maximize expected hazard reduction while satisfying cleaning capacity constraints. UGVs are then deployed to execute cleaning operations accordingly. 

After the cleaning stage, all site hazards evolve according to the dynamic model previously described, which includes both intrinsic growth and spatial influence from nearby locations. Cleaning directly reduces hazard levels before this evolution step, but does not alter the underlying dynamics themselves. Thus, hazard levels can increase again over time even after mitigation, reinforcing the need for proactive and adaptive planning.

Together, these components allow the system to reason under uncertainty, prioritize its limited actions intelligently, and adapt to evolving hazard conditions. By integrating belief modeling, BUCB-based exploration, and task-specific VRPP planning, the framework addresses the dual challenges of dynamic hazard growth and operational constraints.

\subsection{BUCB-Based Sensing Strategy}
We approach the sensing task from a bandit-inspired perspective, where each candidate site is treated as an arm with an uncertain and evolving reward. To balance exploration and exploitation in a principled way, we adopt a BUCB framework. This formulation combines the current hazard estimate with its associated uncertainty to prioritize sensing targets that are either risky or under-explored. It enables the system to (i) prioritize high-risk areas (exploitation), while (ii) actively reduce uncertainty in under-explored sites (exploration). The success of this approach depends on how accurately site-level beliefs capture the evolving hazard state. To that end, we introduce a dynamic belief update mechanism and a distance-aware BUCB score that together guide informed and adaptive sensing decisions. Figure~\ref{fig:sensing_flow} illustrates the operational flow of the sensing phase, from observation processing to score-based routing. For unvisited sites, the system updates the belief state by extrapolating prior estimates, allowing the BUCB score to remain informative even without recent observations.\\

\begin{figure}[htbp]
    \centering
    \resizebox{0.5\linewidth}{!}{%

    \begin{tikzpicture}[node distance=1.5cm]

    \node (start) [startstop] {Start sensing phase};
    \node (receive) [process, below of=start] {Receive observations};
    \node (check) [decision, below of=receive] {Was site observed?};
    
    \node (update) [process, right of=check, xshift=4.3cm] {Time-weighted Bayesian update};
    \node (propagate) [process, below of=check, yshift=-0.5cm] {Extrapolate belief state};
    
    \node (score) [process, below of=propagate] {Compute BUCB score};
    \node (route) [startstop, below of=score] {Use BUCB score in VRPP};
    
    \draw [arrow] (start) -- (receive);
    \draw [arrow] (receive) -- (check);
    \draw [arrow] (check) -- node[above] {Yes} (update);
    \draw [arrow] (update) |- (score);
    \draw [arrow] (check) -- node[left] {No} (propagate);
    \draw [arrow] (propagate) -- (score);
    \draw [arrow] (score) -- (route);

\end{tikzpicture}%
    }
    \caption{BUCB-based sensing workflow. Belief updates and uncertainty propagation feed into BUCB score computation, which guides UAV routing under uncertainty and operational constraints.}
    \label{fig:sensing_flow}
\end{figure}

\subsubsection{Updating Site Beliefs for BUCB}
Accurate and responsive hazard beliefs are essential for computing meaningful BUCB scores at each site. We represent each belief as a Gaussian distribution updated via a time-weighted Bayesian scheme that incorporates recent observations more heavily than older ones. This ensures that the BUCB score reflects the most up-to-date estimate of site risk and uncertainty.

Each site \( i \) maintains a probabilistic belief over its hazard level, modeled as a Gaussian distribution with mean \( \mu_{i,t} \) and variance \( \sigma^2_{i,t} \). These beliefs are updated at each round using a time-weighted Bayesian update that integrates past noisy observations while accounting for their temporal relevance \cite{gelman2013bayesian}. Because the underlying environment is dynamic and hazard levels evolve over time, older measurements may not accurately reflect the current state. To address this, the model assigns exponentially decaying weights to past observations, placing greater emphasis on recent data.

Let \( (y_{i,k}, \tau_{i,k}) \) denote the \( k\)\textsuperscript{th} observation at site \( i \), where \( y_{i,k} \) is the measured hazard and \( \tau_{i,k} \) is the time that the observation was collected. Each observation is weighted by its recency using an exponential decay function,
\begin{equation}
w_{i,k}(t) = \exp\left( -\lambda ( t - \tau_{i,k} ) \right),
\end{equation}
where \( \lambda > 0 \) is a fixed decay rate, and \( t \) is the current time. This scheme reflects the intuition that more recent data is more informative in environments where hazards change over time. Using these weights, the system computes a time-weighted mean of the observations,
\begin{equation}
\bar{y}_{i,t} = \frac{ \sum_k w_{i,k}(t)\, y_{i,k} }{ \sum_k w_{i,k}(t) },
\end{equation}
which represents the current estimated hazard level based on temporally discounted evidence. To modulate the influence of new data, the effective sample size is computed as
\begin{equation}
N_{i,t}^{\text{eff}} = \frac{ \left( \sum_k w_{i,k}(t) \right)^2 }{ \sum_k w_{i,k}^2(t) }.
\end{equation}
This quantity reflects how much reliable information is retained after applying time decay. A smaller \( N_{i,t}^{\text{eff}} \) indicates that only a few recent observations are influential, reducing the confidence of the belief update. Conversely, a larger \( N_{i,t}^{\text{eff}} \) suggests that many past observations remain relevant. We perform a Bayesian update of the belief distribution based on new observations. The variance is updated as
\begin{equation}
\sigma^2_{i,t} = \left( \frac{1}{\sigma^2_{i,t-1}} + \frac{N_{i,t}^{\text{eff}}}{\sigma_\varepsilon^2} \right)^{-1},
\end{equation}
and the mean is updated as
\begin{equation}
\mu_{i,t} = \sigma^2_{i,t} \left( \frac{\mu_{i,t-1}}{\sigma^2_{i,t-1}} + \frac{N_{i,t}^{\text{eff}}\, \bar{y}_{i,t}}{\sigma_\varepsilon^2} \right),
\end{equation}
where \( \sigma_\varepsilon^2 \) denotes the variance of the Gaussian observation noise, which is assumed to be known and constant across all sites and time steps. This parameter reflects the inherent uncertainty in sensor measurements and governs the relative influence of new observations versus prior estimates in the update process.

The resulting formulation effectively balances historical knowledge with recent evidence, producing adaptive and uncertainty-aware estimates that reflect both data quality and temporal dynamics. The complete update procedure is summarized in Algorithm~\ref{alg:twblr}. By emphasizing recent measurements through exponential weighting, the proposed mechanism allows the system to remain responsive to changes in non-stationary environments while still preserving useful prior information.

\begin{algorithm}[htbp]
\caption{Time-Weighted Bayesian Update}\label{alg:twblr}
\begin{algorithmic}[1]
\State \textbf{Input:} Observations $\{(y_{i,k}, \tau_{i,k})\}_{k=1}^{n}$, current time $t$, prior belief $(\mu_{i,t-1}, \sigma^2_{i,t-1})$, noise variance $\sigma_\varepsilon^2$, decay factor $\lambda$
\State Compute time weights: $w_{i,k}(t) = \exp\left( -\lambda (t - \tau_{i,k}) \right)$
\State Compute weighted mean: $\bar{y}_{i,t} = \sum_{k=1}^{n} w_{i,k}(t)\, y_{i,k}$
\State Compute effective sample size: $N_{i,t}^{\text{eff}} = \left( \sum_k w_{i,k}(t) \right)^2 \big/ \sum_k w_{i,k}^2(t)$
\State Compute posterior variance:
\[
\sigma^2_{i,t} = \left( \frac{1}{\sigma^2_{i,t-1}} + \frac{N_{i,t}^{\text{eff}}}{\sigma_\varepsilon^2} \right)^{-1}
\]
\State Compute posterior mean:
\[
\mu_{i,t} = \sigma^2_{i,t} \left( \frac{\mu_{i,t-1}}{\sigma^2_{i,t-1}} + \frac{N_{i,t}^{\text{eff}}\, \bar{y}_{i,t}}{\sigma_\varepsilon^2} \right)
\]
\State \textbf{Output:} Updated belief $(\mu_{i,t}, \sigma^2_{i,t})$
\end{algorithmic}
\end{algorithm}

\subsubsection{Time-Aware Belief Propagation for Unvisited Sites}

While the previous update mechanism ensures accurate beliefs when recent observations are available, many sites may remain unvisited in a given round. To ensure that decisions remain informed, the system continues to generate BUCB scores even in the absence of new observations. To achieve this, we temporally extrapolate site-level beliefs to reflect both increased uncertainty and potential hazard evolution. This involves two key components: inflating the variance to capture growing uncertainty over time, and propagating the mean estimate based on a smoothed hazard gradient.

The variance is increased to reflect reduced confidence as time passes without new observations. Specifically, we apply a linear growth model,
\begin{equation}
\sigma_{i,t+\Delta t}^{2 \mid t} = \min \left[ (1 + \gamma \Delta t)\, \sigma_{i,t}^2,\; \sigma_{\max}^2 \right],
\end{equation}
where \( \gamma > 0 \) is a user-defined inflation rate, \( \Delta t \) is the time since last observation, and \( \sigma_{\max}^2 \) is a saturation threshold that prevents unbounded uncertainty. This mechanism reflects the intuition that belief confidence should degrade as time passes without new evidence.

In parallel, the mean hazard level at each site is extrapolated forward to account for potential changes in unobserved dynamics. This is achieved through a gradient-informed propagation model. When at least two recent observations are available, the instantaneous hazard gradient \( g_{i,t} \) is estimated from recent measurements using exponential smoothing,
\begin{equation}
g_{i,t} = \alpha \cdot \frac{ y_{i,t} - y_{i,t-1} }{ \Delta t } + (1 - \alpha) \cdot g_{i,t-1},
\end{equation}
where \( \alpha \in [0, 1] \) is a smoothing parameter that weights recent versus past gradients. This formulation provides a straightforward yet adaptable approach to capturing local hazard trends from limited data.

We adopt this gradient-based extrapolation due to its simplicity and real-time suitability in settings where explicit hazard models are unavailable. While this approach may be sensitive to noise and trend instability, it remains effective in dynamic, partially observed environments. We further discuss its limitations in \Cref{sec:future_work}.

Using this gradient, the predicted mean is propagated as
\begin{equation}
\mu_{i,t+\Delta t \mid t} = \mu_{i,t} + g_{i,t} \cdot \Delta t,
\end{equation}
yielding an updated estimate of the expected hazard level in the absence of new observations. This propagated estimate informs both the BUCB-based sensing scores and the expected reward used in cleaning decisions. By maintaining belief accuracy over time, the system enables proactive decision-making under partial observability.

\subsubsection{BUCB Score Computation}
Once site beliefs are updated through either recent observations or temporal extrapolation, the system uses them to assign BUCB scores that guide sensing decisions. These scores reflect both the estimated hazard level and the associated uncertainty, allowing the system to balance exploration and exploitation in site selection.

In our formulation, each site is modeled as a bandit arm characterized by a belief distribution over its latent hazard level. This belief comprises a mean estimate \( \mu_{i,t} \) and an associated uncertainty \( \sigma_{i,t}^2 \). To prioritize sensing targets, we compute a BUCB score that balances exploitation and exploration. \changed{For each round $t$, we assign each site $i$ a sensing priority score based on its current belief. Let $S_{i,t}$ denote the raw BUCB score used to rank candidate sensing sites before accounting for routing considerations.}
\begin{equation}\label{eq:raw_bucb}
{S}_{i,t} = \underbrace{\mu_{i,t}}_{\text{exploitation}} + \beta \underbrace{\sqrt{ \sigma_{i,t}^2 }}_{\text{exploration}}.
\end{equation}
\changed{where $\mu_{i,t}$ is the posterior mean hazard estimate at site $i$, $\sigma_{i,t}$ is the posterior standard deviation, and $\beta \ge 0$ controls the exploration-exploitation trade-off.} The mean estimate \( \mu_{i,t} \) reflects the expected hazard level and supports exploitation, while the uncertainty term \( \sigma_{i,t}\) promotes exploration of less-certain sites. Larger values of \( \beta \) encourage the system to explore more aggressively, while smaller values favor exploitation of high-confidence, high-risk areas. This tunable parameter allows the strategy to adapt to different operational priorities or risk preferences, and is selected by the operator based on mission-specific priorities and risk tolerance. 

The BUCB score integrates both the estimated hazard level and its associated uncertainty, making full use of the updated site beliefs to prioritize sensing targets. Sites with high estimated hazard or large uncertainty are favored, resulting in a dynamic tradeoff between information gathering and immediate intervention. This scoring formulation serves as the basis for site prioritization during the sensing phase. While the raw BUCB values capture intrinsic site utility under uncertainty, additional adjustments are introduced in the routing strategy to account for spatial cost.

\subsection{Task-Specific Routing Strategies}
We formulate both the sensing and cleaning phases as instances of the VRPP, enabling selective site visits under operational constraints. While the core structure of the routing formulation remains consistent, task-specific objectives and constraints lead to different implementations in the sensing and cleaning phases. We first introduce the general VRPP model, then describe how it is specialized for each task.

\subsubsection{General VRPP Formulation}
To enable selective site visits under limited vehicle resources, we adopt a vehicle routing framework for both the sensing and cleaning phases. Each task is formulated as a VRPP, which seeks to maximize the total visit value collected from selected sites without requiring full coverage of all sites. \changed{\Cref{fig:vrp_conops} shows the vehicle routing environment considered in this formulation.} This selective formulation aligns well with our setting, where visiting every site is infeasible and target selection must adapt dynamically based on evolving hazard estimates and operational constraints.

\begin{figure}[h!]
    \centering
    \includegraphics[width=0.3\linewidth]{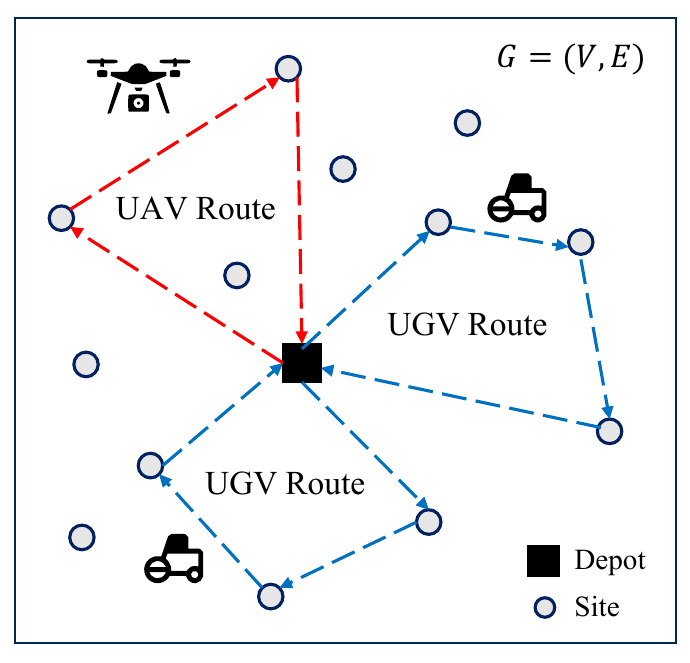}
    \caption{\changed{Vehicle routing environment for the VRPP formulation, showing a depot and candidate sites, where red indicates UAV sensing routes and blue indicates UGV cleaning routes.}}
    \label{fig:vrp_conops}
\end{figure}

The environment is represented as a graph \( G = (V, E) \), where \( V = \{0\} \cup S \) consists of the depot node \( 0 \) and a set of candidate sites \( S \), and \( E \subseteq V \times V \) is the set of edges. Each edge \( (i, j) \in E \) has an associated distance \( d_{ij} \). The vehicle fleet is denoted by \( M \). Although the formulation uses a generic notation for simplicity, our implementation assigns distinct fleets for sensing and cleaning to reflect their different operational roles. Each site \( i \in S \) has an associated visit value \( \pi_i \), representing the utility of visiting that site, with BUCB scores in the sensing phase and expected cleaning rewards in the cleaning phase. In cleaning, each site also has a demand \( \delta_i \) that indicates the required cleaning load. We define binary variables \( x_{im} \) and \( z_{ijm} \) to represent whether vehicle \( m \) visits site \( i \) and traverses edge \( (i, j) \), respectively. The auxiliary variable \( u_{im} \) is used to eliminate subtours.

The general VRPP formulation is given in \eqref{eq:vrp_formulation}.
\begin{subequations}\label{eq:vrp_formulation}
\begin{align}
\max_{x,z}\;
    &\sum_{m\in M}\sum_{i\in V\setminus\{0\}}\pi_i\,x_{im}
      -c\sum_{m\in M}\sum_{(i,j)\in E} d_{ij}\,z_{ijm}
      \label{eq:vrp_obj} \\[6pt]
\text{s.t. }\;
    &\sum_{j:(i,j)\in E} z_{ijm}
     =\sum_{j:(j,i)\in E} z_{jim}=x_{im},
      &&\forall i\in V \setminus \{0\},\;\forall m \in M,
      \label{eq:vrp_flow_conservation} \\[6pt]
    &\sum_{m\in M}x_{im}\le 1,
      &&\forall i\in V \setminus \{0\}, \quad\text{(sensing only)}
      \label{eq:vrp_sensing_unique_visit} \\[6pt]
    &\sum_{(i,j)\in E} d_{ij}\,z_{ijm}\le D_m,
      &&\forall m \in M, \quad\text{(sensing only)}
      \label{eq:vrp_sensing_distance_limit} \\[6pt]
    &\sum_{i\neq0}\delta_i\,x_{im}\le Q_m,
      &&\forall m \in M, \quad\text{(cleaning only)}
      \label{eq:vrp_cleaning_capacity} \\[6pt]
    &\sum_{j\neq0}z_{0jm}=1,\;
      \sum_{i\neq0}z_{i0m}=1,
      &&\forall m \in M,
      \label{eq:vrp_depot_constraints} \\[6pt]
    &u_{im}-u_{jm}+|S|\,z_{ijm}\le |S|-1,
      &&\forall(i,j)\in E,\;i,j\neq0,\;\forall m \in M,
      \label{eq:vrp_subtour_elimination} \\[6pt]
    &x_{im}\in\{0,1\},\;z_{ijm}\in\{0,1\},\;
      1\le u_{im}\le|S|.
      \label{eq:vrp_variable_domains}
\end{align}
\end{subequations}

The objective function in \Cref{eq:vrp_obj} maximizes the total visit value from visiting selected sites while subtracting the total travel cost. 
Routes are primarily determined by the visit values associated with each site, which reflect sensing utility or cleaning reward depending on the phase. 
The inclusion of travel cost serves to discourage inefficient or redundant routing, promoting more compact and cost-effective tours without compromising the overall objective. We seek to minimize the objective subject to a set of constraints.
Flow conservation is enforced by \Cref{eq:vrp_flow_conservation}, ensuring that each visited site has exactly one incoming and one outgoing edge. 
Sensing-specific constraints include \Cref{eq:vrp_sensing_unique_visit}, which restricts each site to a single visit across all sensing UAVs. This encourages broader coverage and avoids redundant observations of the same site. In contrast, the cleaning phase does not impose this restriction, allowing multiple UGVs to service the same site within a round if the hazard level is high enough to require more than one cleaning effort. 
\Cref{eq:vrp_sensing_distance_limit} imposes an additional sensing-specific constraint, limiting the total travel distance of each UAV route to its maximum range \( D_m \).
For cleaning, \Cref{eq:vrp_cleaning_capacity} limits the total amount of hazard cleaned on each route to the vehicle’s capacity \( Q_m \). These per-vehicle parameters enable support for heterogeneous fleets, allowing vehicles to differ in sensing range or cleaning capacity. 
Depot entry and return are guaranteed by \Cref{eq:vrp_depot_constraints}, which requires all routes to begin and end at the depot node. 
To eliminate subtours and ensure route connectivity, \Cref{eq:vrp_subtour_elimination} applies the Miller-Tucker-Zemlin (MTZ) formulation \cite{10.1145/321043.321046}. 
Finally, \Cref{eq:vrp_variable_domains} imposes domain constraints on the decision variables.
The formulation accommodates both sensing and cleaning tasks through flexible routing, while respecting task-specific reward structures and operational constraints. Table~\ref{tab:vrpp_comparison} provides a summary of how the general VRPP formulation is applied differently for sensing and cleaning.

\begin{table}[htbp]
\centering
\caption{Comparison of VRPP formulations for sensing and cleaning phases}
\label{tab:vrpp_comparison}
\begin{tabular}{lcc}
\toprule
\textbf{Component} & \textbf{Sensing phase (UAV)} & \textbf{Cleaning phase (UGV)} \\
\midrule
Objective visit value         & BUCB score  & Expected hazard reduction \\
Visit constraint                    & At most one vehicle per site      & Multiple visits allowed \\
Route budget constraint             & Maximum distance    & ---\tablefootnote{A route budget constraint is not explicitly applied in the cleaning phase since the capacity constraint is typically more restrictive. Route efficiency is still encouraged indirectly through the objective formulation.} \\
Capacity constraint        & Not applied                       & Total cleaning capacity \\
Per-site demand       & Not defined                       & Required (cleaning load) \\
Planning goal                  & Information gain under budget     & Hazard mitigation within capacity \\
\bottomrule
\end{tabular}
\end{table}

\subsubsection{Sensing Strategy in Routing}
In the sensing phase, each site's visit value \( \pi_i^{\text{sense}} \) in the VRPP is derived from its BUCB score, as defined in \Cref{eq:raw_bucb}, which reflects both the estimated hazard and associated uncertainty. These scores are updated at every decision round based on the latest site-level beliefs, allowing the system to adapt dynamically to new observations. However, directly using these scores can lead to routing bias: since vehicle routing naturally favors nearby sites to reduce travel cost, distant but highly informative sites are under-selected.

\changed{To mitigate this bias while preserving the relative ordering induced by the BUCB scores, we introduce a distance-aware adjustment that rescales each site's sensing utility based on its spatial cost.}
\begin{equation}
\tilde{S}_{i,t} = \frac{ {S}_{i,t} }{ 1 + \kappa d_i },
\end{equation}
where \( d_i \) is the distance from the depot to site \( i \), and \( \kappa > 0 \) governs the strength of the spatial penalty. 
\changed{This multiplicative penalty softly discounts distant sites without eliminating them outright, allowing highly hazardous or uncertain locations to remain competitive when their informational value outweighs travel cost. The adjusted score \( \tilde{S}_{i,t} \) is then used as the final visit value used by the sensing VRPP, i.e., \( \pi_i^{\text{sense}} = \tilde{S}_{i,t} \).}

By integrating both hazard-driven informational value and spatial cost, this strategy enables UAVs to prioritize high-risk or uncertain sites even when they are far from the depot, correcting for the routing bias that would otherwise favor only nearby locations.

\subsubsection{Cleaning Strategy in Routing}

In the cleaning phase, route planning is updated at each round based on the current belief about the site hazard level. We define an adjusted hazard estimate that accounts for both the expected hazard and its associated uncertainty to prioritize sites.

The actual amount of hazard that can be removed from a site is constrained by a per-site cleaning capacity,
\(R_{i,t} = \min\left( {\mu}_{i,t},\, Q_{\text{unit}} \right).\)
This limit reflects real-world restrictions on the amount of cleaning action that can be performed at a single location during a single visit. For example, a UGV may have safety protocols that cap the amount of hazard it is allowed to handle before returning to base \cite{robotics10020078, schwaiger2024ugvcbrnunmannedgroundvehicle}. This constraint ensures that the system models feasible on-site interventions and avoids unrealistic one-shot cleaning assumptions.

The estimated cleaning reward is given by
\( \pi_i^{\text{clean}} = \mu_{i,t} \cdot R_{i,t},\)
representing the expected hazard reduction if the site is visited and cleaned. \changed{The product form captures the intuition that effective cleaning depends both on the estimated hazard severity at a site and on the amount of mitigation effort that can be applied during the visit, thereby representing the expected hazard reduction while retaining sensitivity to the belief estimate \( \mu_{i,t} \).} As a result, when two sites yield the same deliverable quantity, the system prioritizes the one with greater estimated risk. The corresponding cleaning demand for each site is defined as \(\delta_i = R_{i,t}.\)

Each vehicle is also subject to a total cleaning capacity constraint, as given in \Cref{eq:vrp_cleaning_capacity}. This route-level constraint reflects cumulative resource limits such as tank volume or waste storage, ensuring that the total cleaning load assigned along each UGV’s route remains feasible within its overall capacity. Together, these constraints ensure that site-level feasibility and overall resource limits are both respected during the cleaning phase.

Finally, we account for residual hazard cases, where the belief suggests that a site has been fully cleaned (\(\mu_{i,t} = 0\)) but the UGV reports that some hazard remains after cleaning the site. While the cleaning UGV may not quantify the exact remaining amount, we assume it can determine whether the hazard has been fully eliminated. This discrepancy suggests the system may have been overconfident in its previous estimate. Such sites are likely to be excluded from future sensing due to their low estimated hazard, which delays hazard removal and reduces mission effectiveness. To address this, we increase the estimated uncertainty at these sites to promote future sensing and reduce the risk of overlooked residual hazards. Specifically, we apply an uncertainty boost \( \zeta \) to the variance estimate when residual hazard exists despite the belief indicating successful cleaning. 
\begin{equation} \label{eq:update_uncertainty}
\sigma_{i,t}^2 \leftarrow \min\left( \sigma_{i,t}^2 + \zeta,\, \sigma_{\max}^2 \right)
\qquad \text{if } \mu_{i,t} = 0 \;\text{and}\; H_{i,t} > 0.
\end{equation}
While our framework currently treats UGVs purely as cleaning agents, it is worth noting that many UGV platforms can also carry sensing payloads. If equipped with such capabilities, their observations could be integrated into the sensing module and associated VRPP rounds in future extensions of this work.

\subsection{End-to-End Strategy}

Our proposed method integrates the sensing and cleaning phases into a unified round-based control loop. At each round, the system evaluates all candidate sites using BUCB scores, which reflect both estimated hazard levels and associated uncertainty. These scores define the visit values in a sensing VRPP, which is solved to determine efficient UAV routes for sensing under range constraints. After collecting noisy observations from the selected sites, the belief model is updated using a time-weighted Bayesian update, while unvisited sites undergo extrapolation and variance inflation.

Using the updated beliefs, the system estimates expected cleaning rewards and demand levels at each site. A second VRPP is then solved to assign UGV cleaning routes that maximize expected hazard reduction subject to vehicle capacity constraints. If a site remains hazardous despite being predicted clean, its belief uncertainty is increased to encourage future sensing. All hazard levels evolve according to the underlying environmental dynamics. This process repeats until all sites fall below a hazard threshold or the maximum number of rounds is reached. The complete decision procedure is summarized in Algorithm~\ref{alg:end2end}.

\begin{algorithm}[htbp]
\caption{BUCB-Based Vehicle Dispatch and Routing Strategy}\label{alg:end2end}
\begin{algorithmic}[1]
\State \textbf{Input:} Site set $S$, sensing vehicle fleet $M^s$, cleaning vehicle fleet $M^c$, max rounds $T$, parameters
\State Initialize true hazard $H_i$ for all $i \in S$
\State Initialize belief $(\mu_i, \sigma^2_i)$ for all $i \in S$
\For{$t = 1$ to $T$}
    \State Compute BUCB score $S_{i,t} = \mu_{i,t} + \beta \sqrt{\sigma_{i,t}^2}$ for all $i \in S$ (\Cref{eq:raw_bucb})
    \State Adjust scores using distance-aware penalty
    \State Use fleet $M^s$ to solve the sensing-phase VRPP with distance-aware BUCB-based visit values
    \State Sample noisy observations $y_{i,t} \sim \mathcal{N}(H_{i,t}, \sigma^2_\varepsilon)$ for selected sites
    \State Update beliefs for observed sites via time-weighted Bayesian update
    \State Propagate mean and inflate variance for unobserved sites
    \State Compute cleaning reward: 
    $R_{i,t} = \min({\mu}_{i,t}, Q_\text{unit})$,
    $\pi_{i,t}^\text{clean} = \mu_{i,t} \cdot R_{i,t}$, 
    $\delta_{i,t} = R_{i,t}$
    \State Use fleet $M^c$ to solve the cleaning-phase VRPP with hazard-based visit values
    \State Apply hazard reduction and belief update for cleaned sites
    \For{each $i \in S$}
        \If{$\mu_{i,t} = 0$ \textbf{and} $H_{i,t} > 0$}
            \State Increase uncertainty: $\sigma^2_{i,t} \leftarrow \min(\sigma^2_{i,t} + \zeta, \sigma^2_{\max})$ (\Cref{eq:update_uncertainty})
        \EndIf
    \EndFor

    \If{$H_{i,t} = 0$ for all $i$}
        \State \textbf{break}
    \EndIf
    \State Simulate hazard growth and spatial propagation
\EndFor
\State \textbf{Output:} Fully mitigated hazard states and complete UAV/UGV routing history
\end{algorithmic}
\end{algorithm}

\section{Experimental Results}

\subsection{Simulation Setup and Evaluation Metrics}
We present the simulation setup, including key parameters used throughout the experiments. We then define the evaluation metrics used to assess hazard mitigation performance and the accuracy of belief estimation.

\subsubsection{Simulation Environment}
The simulation operates in a discrete-time, round-based structure, alternating between sensing and cleaning stages. Each round begins with UAV-based sensing to collect noisy observations from selected sites, followed by UGV-based cleaning at prioritized locations. Key parameters used throughout the simulation are summarized in \Cref{tab:sim_env_params} and \Cref{tab:sim_vehicle_params}.

Hazard levels at each site evolve over time following a logistic growth model with added spatial diffusion from nearby sites, as detailed in Section~\ref{sec:hazard_dynamics}. Initial hazard levels are sampled uniformly from \([0, 100]\), and intrinsic growth rates \(\rho_i\) are drawn from the uniform distribution \(\mathcal{U}(0.0, 0.1)\). If all hazards are cleared, the simulation terminates early; otherwise, it proceeds for a maximum of 50 rounds. As shown in \Cref{fig:hazard_dynamics}, hazard levels tend to increase over time in the absence of intervention, reinforcing the necessity of proactive cleaning operations. This example follows the same simulation settings used in our main experiments, with hazards evolving through logistic growth and spatial diffusion.

\begin{figure}[h!]
    \centering
    \begin{subfigure}[b]{0.45\linewidth}
        \centering
        \includegraphics[width=\linewidth]{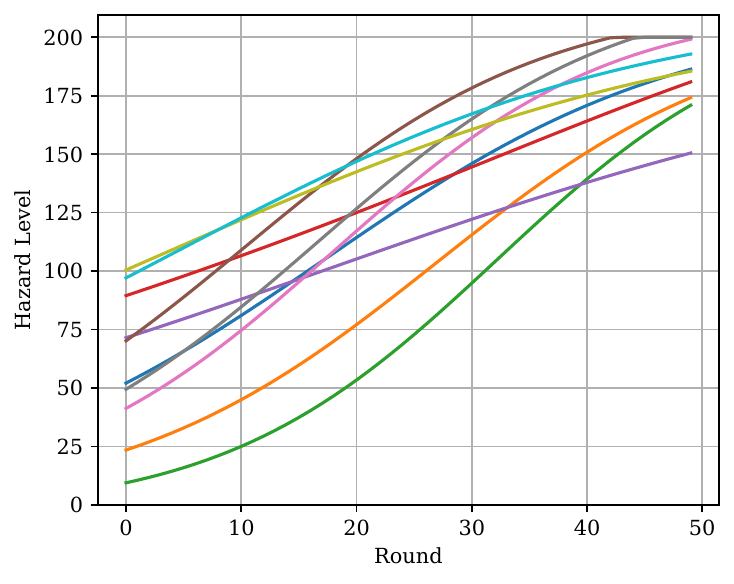}
        \caption{Hazard level changes.}
        \label{fig:hazard_dynamics_left}
    \end{subfigure}
    \hfill
    \begin{subfigure}[b]{0.45\linewidth}
        \centering
        \includegraphics[width=\linewidth]{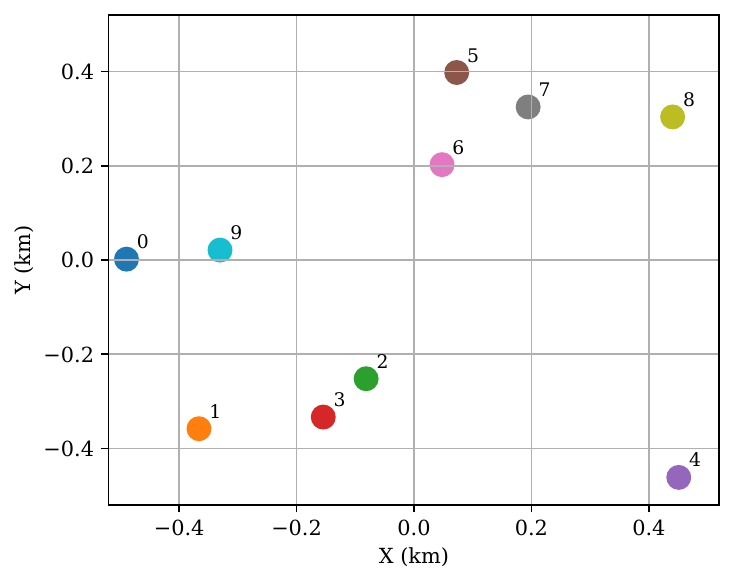}
        \caption{Physical locations of the same sites.}
        \label{fig:hazard_dynamics_right}
    \end{subfigure}
    
    \caption{\changed{Hazard dynamics simulation for 10 sites: (a) evolution without cleaning intervention; (b) spatial layout of the sites.}}
    \label{fig:hazard_dynamics}
\end{figure}

Both the sensing and cleaning VRPPs are solved using PyVRP, an open-source solver based on a hybrid genetic algorithm \cite{pyvrp}. Site coordinates are sampled uniformly from a square region with x- and y-coordinates in $[-0.5, 0.5]$ km map centered at the origin, where the depot is located. The number of sensing UAVs and cleaning UGVs is fixed during the simulation. \changed{Each VRPP instance is solved with a fixed time limit of 30 seconds, ensuring bounded runtime per decision round while maintaining high-quality routing solutions in both sensing and cleaning phases.} Experiments were conducted on a machine equipped with a 13\textsuperscript{th} Gen Intel\textregistered{} Core\texttrademark{} i7-1360P CPU. 

\begin{table}[h]
\centering
\caption{parameters for environment dynamics and BUCB-based decision logic.}
\label{tab:sim_env_params}
\begin{tabular}{llc}
\toprule
\textbf{parameter} & \textbf{Description} & \textbf{Value} \\
\midrule
$\rho_i$ & Intrinsic growth rate & $\sim \mathcal{U}(0.0,\;0.1)$ \\
$K$ & Maximum Hazard & 200 \\
$\phi$ & Spatial influence coefficient & 0.01 \\
$\sigma_\varepsilon$ & Sensor noise standard deviation & 5.0 \\
$\mu_{i,0}$ & Prior belief mean & 0.0 \\
$\sigma^2_{i,0}$ & Prior belief variance & 100.0 \\
$\lambda$ & Time-decay rate in Bayesian update & 0.5 \\
$\gamma$ & Variance inflation rate & 0.5 \\
$\alpha$ & Gradient smoothing weight & 0.3 \\
$\beta$ & BUCB exploration coefficient & 20.0 \\
$\zeta$ & Uncertainty boost after failed cleaning & 100.0 \\
\bottomrule
\end{tabular}
\end{table}

\begin{table}[h]
\centering
\caption{Vehicle routing parameters used in the simulation.}
\label{tab:sim_vehicle_params}
\begin{tabular}{llc}
\toprule
\textbf{parameter} & \textbf{Description} & \textbf{Value} \\
\midrule
Map size & Spatial extent of the environment & $[-0.5,\;0.5] \times [-0.5,\;0.5]$ km \\
$D_m$ & Max route distance per UAV & 1.5 km \\
$Q_{\text{unit}}$ & Max cleaning per site visit & 25 \\
$Q_m$ & Cleaning capacity per UGV & 100 \\
$c$ & Unit travel cost & 1.0 per km \\
$\kappa$ & Distance penalty coefficient & 0.1 per km \\
\bottomrule
\end{tabular}
\end{table}

\subsubsection{Evaluation Metrics}
\label{sec:evaluation_metrics}
The following metrics are used to evaluate the effectiveness of the simulation. 
\begin{itemize}
    \item {Termination Rounds}: Number of simulations until a stopping criteria is reached, defined as
    \begin{equation}
        T_{\text{end}} = \min \left( \left\{ t \leq T : H_{i,t} = 0,\; \forall i \in \{1, \dots, N\} \right\} \cup \{T\} \right)
    \end{equation}
    where \(T\) is the maximum allowed number of rounds.

    \item {Cumulative Hazard}: The total accumulated hazard across all sites and rounds,
    \( \sum_{t=1}^{T} \sum_{i=1}^{N} H_{i,t} \)
    which aligns with the objective function \(\mathcal{J}\) in \Cref{sec:problem_statement}.

    \item {Hazard Mitigation Rate}: Average amount of hazard removed per round, \( \frac{1}{T} \sum_{t=1}^{T} C_t \)
    where \(C_t\) is the total hazard removed at round \(t\).

    \item {Final Mean Absolute Error (MAE)}: MAE between estimated and true hazard at the final round \(T\),
    \begin{equation}
        \text{Final MAE} = \frac{1}{N} \sum_{i=1}^{N} \left\lvert \mu_{i,T_{\text{end}}} - H_{i,T_{\text{end}}} \right\rvert
    \end{equation}
    measuring belief accuracy at termination.

\end{itemize}

\subsection{Performance Evaluation of the Proposed Method}
\label{sec:proposed_method}
To assess the effectiveness of our proposed framework, we conduct simulations under multiple environment scenarios. We evaluate three simulation scenarios, summarized in \Cref{tab:simulation_scenarios}. Each scenario varies in the number of hazard sites, the number of sensing UAVs, and the number of cleaning UGVs. Each scenario is repeated across 100 random seeds to ensure statistical robustness, resulting in a total of 300 simulation runs.

\begin{table}[h]
\centering
\caption{Simulation scenario configurations}
\begin{tabular}{lccc}
\toprule
\textbf{Scenario} & \textbf{Hazard Sites} & \textbf{Sensing UAVs} & \textbf{Cleaning UGVs} \\
\midrule
Scenario 1 & 20 & 2 & 2 \\
Scenario 2 & 50 & 2 & 2 \\
Scenario 3 & 50 & 2 & 3 \\
\bottomrule
\end{tabular}
\label{tab:simulation_scenarios}
\end{table}

The analysis proceeds in \changed{three} stages. We first present overall performance across all scenarios, focusing on aggregate statistics and temporal trends. The following subsections then examine a representative single case to illustrate routing behavior and site-level outcomes. \changed{Finally, we conduct a sensitivity analysis to assess the robustness of the proposed framework under varying hazard growth rates.}

\subsubsection{Overall Performance}
\Cref{tab:key_metrics_summary} presents the overall performance metrics across the three simulation scenarios. Each metric is defined formally in \Cref{sec:evaluation_metrics}. \changed{All scenarios achieve a 100\% success rate, with full hazard clearance occurring before the specified maximum round limit in all tested cases.} \changed{The observed 100\% hazard clearance reflects the tested experimental settings and operational assumptions, rather than a guarantee under arbitrary hazard dynamics or deployment conditions.}

In the smallest setting (Scenario 1), full mitigation is achieved \changed{in fewer than} 7 rounds on average. As expected, increasing the number of sites while keeping the fleet size fixed leads to longer simulations; Scenario 2 requires approximately 20 rounds on average. Notably, despite the increased problem size, Scenario 2 exhibits a higher hazard cleaning rate than Scenario 1. This efficiency gain arises because larger environments reduce the relative duration of early exploration and late-stage low-hazard cleanup, allowing more rounds to be spent in effective full-capacity cleaning. \changed{In Scenario~2, more rounds are spent cleaning sites with substantial remaining hazard at near-capacity operation, while smaller environments enter low-impact cleanup phases earlier. Consequently, the average hazard reduction per round increases, yielding a higher cleaning rate.} The addition of a third UGV (Scenario 3) reduces the average number of rounds to around 12, indicating that the system effectively exploits added cleaning capacity to accelerate mitigation.

Cumulative hazard increases significantly with both the number of sites and the simulation duration. Comparing Scenarios 2 and 3 shows that adding a third UGV substantially reduces the cumulative hazard, from over 26,000 to approximately 14,800. This suggests that increasing cleaning capacity can significantly reduce overall hazard accumulation in extensive environments. However, the improvement is not linear, reflecting diminishing returns and inherent physical constraints on the amount of hazard that can be mitigated per round. Final MAE remains low across all scenarios, indicating robust belief tracking. The lowest error is in Scenario 2, due to the increased sensing and refinement resulting from a longer simulation. 

\begin{table}[htbp]
\centering
\caption{Summary of key performance metrics across experimental scenarios.}
\label{tab:key_metrics_summary}
\begin{tabular}{lccccc}
\toprule
\textbf{Scenario} &
\textbf{Termination Rounds} &
\textbf{Cumulative Hazard}&
\textbf{Hazard Cleaning Rate} &
\textbf{Final MAE}\\
\midrule
Scenario 1 &
6.82 ± 1.03 &
3355.02 ± 933.25 &
161.04 ± 14.63 &
1.17 ± 1.06 \\
Scenario 2 &
20.02 ± 2.33 &
26150.08 ± 5186.27 &
187.73 ± 7.53 &
0.40 ± 0.39 \\
Scenario 3 &
12.07 ± 0.99 &
14863.46 ± 2420.60 &
260.52 ± 13.10 &
0.51 ± 0.63 \\
\bottomrule
\end{tabular}
\end{table}

\begin{figure}[htbp]
    \centering
    \begin{subfigure}[b]{0.48\linewidth}
        \centering
        \includegraphics[width=\linewidth]{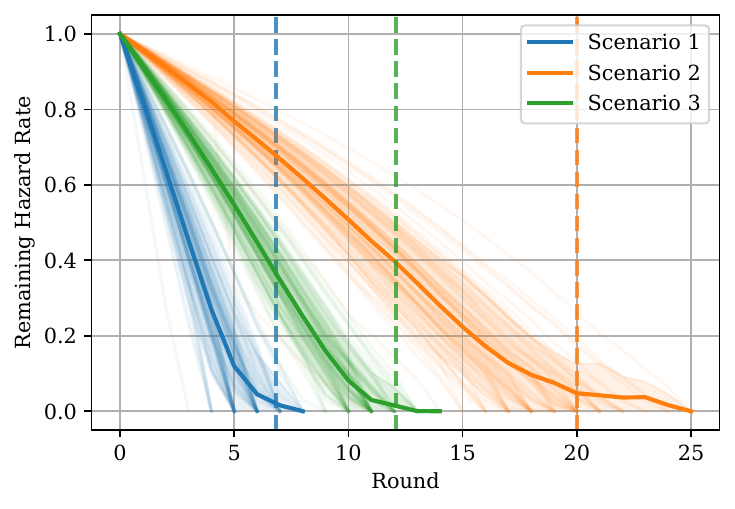}
        \caption{Remaining hazard rate over time.}
        \label{fig:remaining_hazard_curve}
    \end{subfigure}
    \hfill
    \begin{subfigure}[b]{0.48\linewidth}
        \centering
        \includegraphics[width=\linewidth]{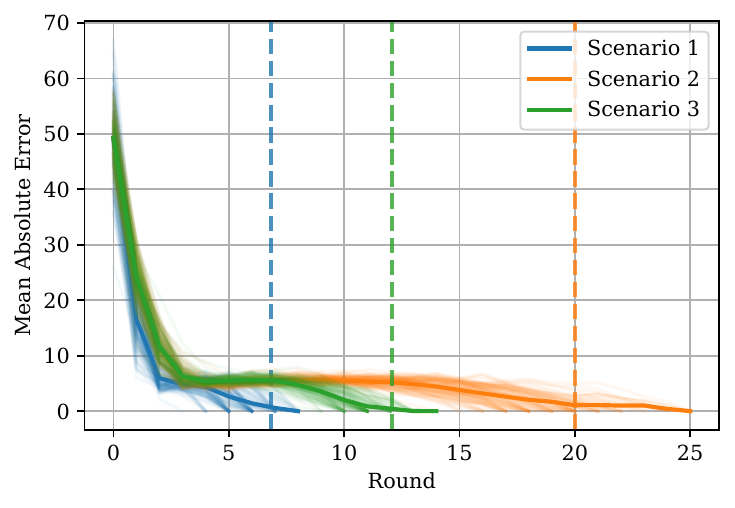}
        \caption{MAE over time.}
        \label{fig:mae_curve}
    \end{subfigure}
    \caption{Main performance metrics across scenarios. \changed{Dashed lines show average termination rounds; bold lines indicate means, with transparent lines showing runs within the 10th–90th percentiles.}}
    \label{fig:hazard_mae_split}
    \end{figure}

\Cref{fig:hazard_mae_split} visualizes the round-wise evolution of two key performance metrics: remaining hazard and BUCB-based belief accuracy. At each round, the remaining hazard is computed as the mean hazard level across all sites, with results shown across 100 independent runs per scenario. As shown in \Cref{fig:remaining_hazard_curve}, hazard levels decrease consistently across rounds, with more rapid mitigation observed in scenarios with additional cleaning UGVs and fewer hazard sites. \Cref{fig:mae_curve} displays the MAE between estimated and true hazard values at each round. MAE trends downward sharply after the initial few rounds and continues to decline more gradually as the simulation approaches termination, indicating progressive improvement in hazard estimation over time.

\begin{figure}[htbp]
    \centering
    \begin{subfigure}[t]{0.57\textwidth}
        \includegraphics[width=\linewidth]{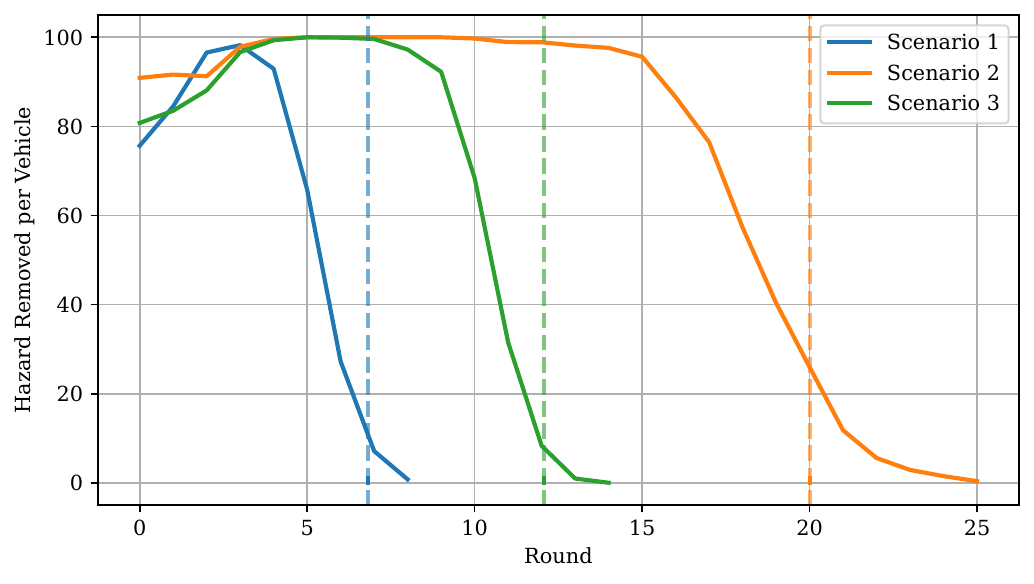}
        \caption{Hazard removed per-UGV over time.}
        \label{fig:hazard_curve}
    \end{subfigure}
    \hspace{1em}
    \begin{subfigure}[t]{0.278\textwidth}
        \includegraphics[width=\linewidth]{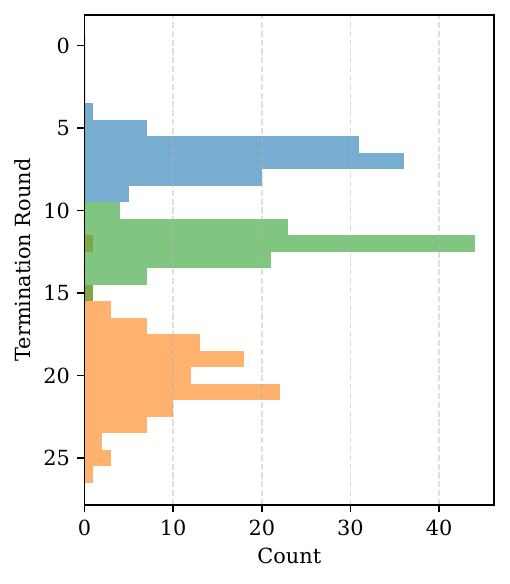}
        \caption{Histogram of termination rounds across runs}
        \label{fig:termination_hist}
    \end{subfigure}

    \caption{
        Per-UGV hazard mitigation efficiency and termination behavior across simulation scenarios. 
    }
    \label{fig:hazard_hist_combined}
\end{figure}

\Cref{fig:hazard_hist_combined} summarizes per-UGV hazard mitigation efficiency and termination round distribution across three simulation scenarios. All three scenarios exhibit qualitatively similar mitigation dynamics, validating consistent system behavior across scales. In \Cref{fig:hazard_curve}, each curve represents the average hazard removed per cleaning UGV at each round, calculated over 100 independent runs. Since the total number of rounds varies, later rounds include fewer cases and are thus less representative. To aid interpretation, the average termination round for each scenario is marked with a dashed vertical line, and the termination histogram is shown in \Cref{fig:termination_hist}.

As expected, the first few rounds (typically up to round 3) are dominated by exploration, during which UGVs operate below full capacity due to high uncertainty in hazard estimates. This is followed by a stable phase where UGVs consistently operate near full capacity, removing a steady amount of hazard each round. The pattern is most clearly observed in the 50-site scenarios, where the larger scale enables clearer phase separation. In contrast, Scenario 1 shows no sustained flat region due to fewer sites and shorter duration. Near the end of each run, the mitigation per-UGV declines again as site hazards are mostly depleted. 

\begin{figure}[htbp]
    \centering
    \includegraphics[width=0.5\linewidth]{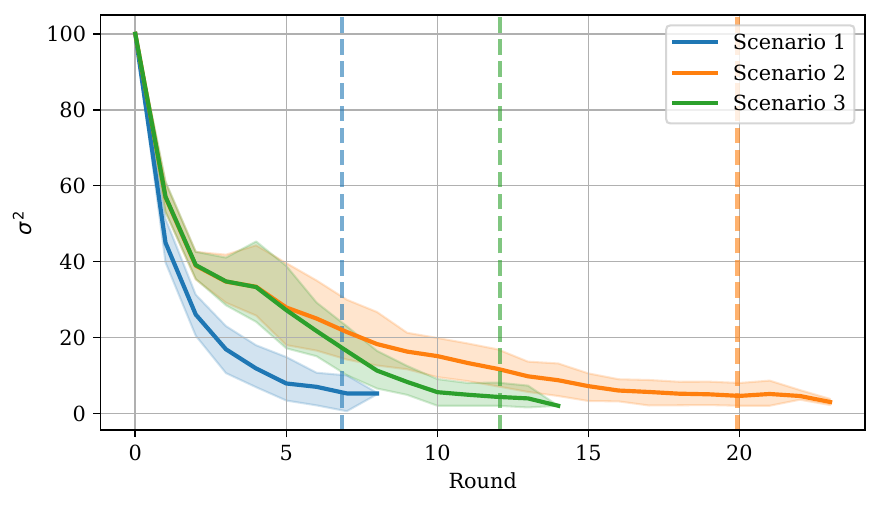}
    \caption{Convergence of posterior variance over rounds, averaged across sites. Solid lines show means across \changed{runs, with shaded regions indicating the 10th–90th percentile range.}}
    \label{fig:sigma2_conv}
\end{figure}

\Cref{fig:sigma2_conv} illustrates the evolution of the posterior variance $\sigma^2$ across rounds, averaged over all sites and 100 simulation runs per scenario. Since each scenario terminates at a different number of rounds, vertical dashed lines are included to mark the average termination round, enabling fair interpretation. Across all scenarios, posterior variance decreases rapidly in the initial few rounds as the system collects informative observations, and converges near zero by the final round. This consistent downward trend confirms the effectiveness of the belief update mechanism in reducing uncertainty over time.

\begin{figure}[htbp]
    \centering
    \includegraphics[width=0.5\linewidth]{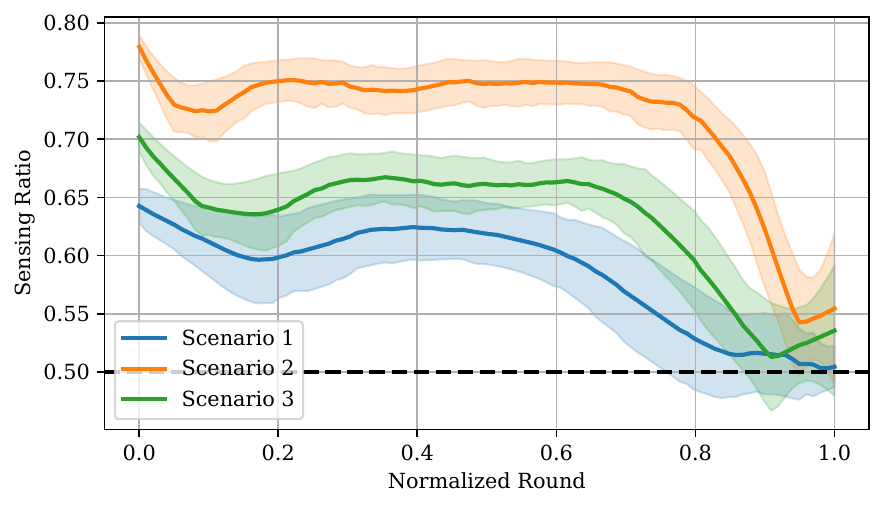}
    \caption{
    Mean sensing ratio over normalized rounds across scenarios. \changed{Curves show average sensing allocation; shaded areas indicate variability, with a dashed line marking balanced allocation.}}
    \label{fig:sensing_ratio}
\end{figure}

\Cref{fig:sensing_ratio} illustrates the mean sensing ratio over normalized simulation rounds across the three scenarios. The sensing ratio at each round is defined as the number of sites visited by sensing UAVs divided by the total number of distinct sites visited by both UAVs and UGVs. The horizontal dashed line at 0.5 marks a balanced allocation between exploration and exploitation, corresponding to an equal number of visited sites allocated to sensing and cleaning within each round. Each curve represents the average trajectory across 100 runs, with shaded bands indicating the standard deviation. Rounds are normalized by each run's total duration to enable cross-setting comparison.

The overall trend is consistent and interpretable. During the initial rounds, the sensing ratio is high as UAVs are prioritized to gather information in the face of high uncertainty. As beliefs become more accurate, the system shifts toward increased cleaning, stabilizing the sensing ratio. In the middle of the trajectory, the ratio remains relatively steady during the phase of full-capacity UGV operation. Toward the final rounds, the ratio declines significantly. This occurs because in later rounds, most remaining sites have minimal residual hazard, often less than the per-visit removal limit \(Q_\text{unit}\). Consequently, UGVs visit more sites to fully utilize their cleaning capacity, which increases the total number of cleaning targets and reduces the sensing ratio. These dynamics confirm that the framework allocates resources adaptively in response to evolving hazard conditions.

\subsubsection{Single-Case Visualization}
To complement the aggregate trends, we examine a representative single run from Scenario 2. This case study provides a detailed view of how the system allocates sensing and cleaning tasks over time and space. We focus on key rounds to highlight how the hazard belief, vehicle routing, and mitigation efforts evolve throughout the simulation.

\begin{figure}[htbp]
    \centering
    \includegraphics[width=0.55\linewidth]{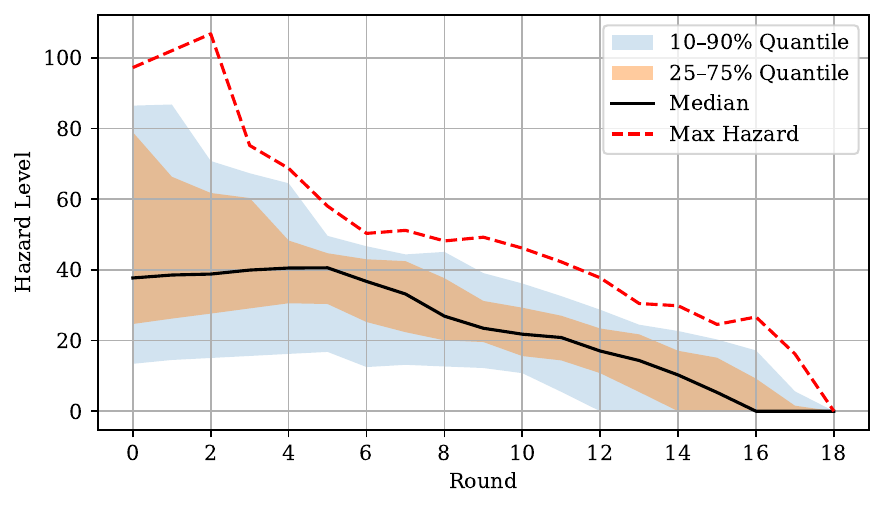}
    \caption{\changed{Temporal trend of true hazard levels over rounds, showing site quantiles, median, and maximum.}}
    \label{fig:single_reduction}
\end{figure}

\Cref{fig:single_reduction} illustrates the temporal evolution of \changed{true} hazard levels across all sites in a representative simulation run. During the first two rounds, sensing does not yet cover all sites, increasing the \changed{true} maximum hazard. A sharp drop in maximum hazard is observed immediately after round 3, following the first full sensing phase and subsequent targeted cleaning of the most hazardous sites. Notably, the median hazard continues to rise until around round 5. This trend reflects the system's strategy of prioritizing high-hazard sites early, which prevents any individual site from reaching the maximum threshold level \( K \). The narrowing shaded region over the initial rounds indicates that site-level hazards gradually converge toward a similar range, reflecting a more uniform reduction in hazard across the environment. After round 5, hazard levels stabilize, and the system maintains a steady hazard reduction.

\begin{figure}[htbp]
    \centering

    \begin{subfigure}[b]{0.325\textwidth}
        \includegraphics[width=\textwidth]{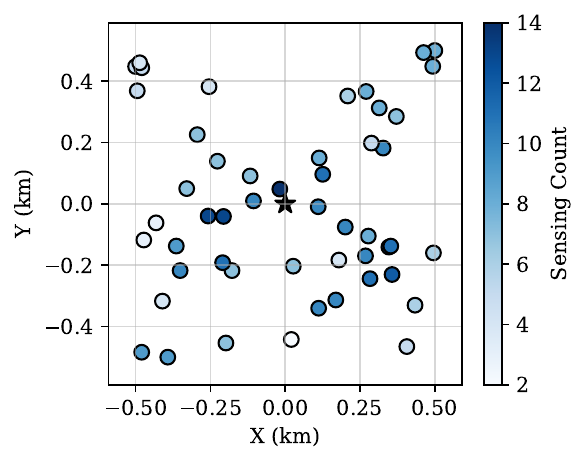}
        \caption{Sensing activity map}
        \label{fig:sensing_activity_map}
    \end{subfigure}
    \hfill
    \begin{subfigure}[b]{0.325\textwidth}
        \includegraphics[width=\textwidth]{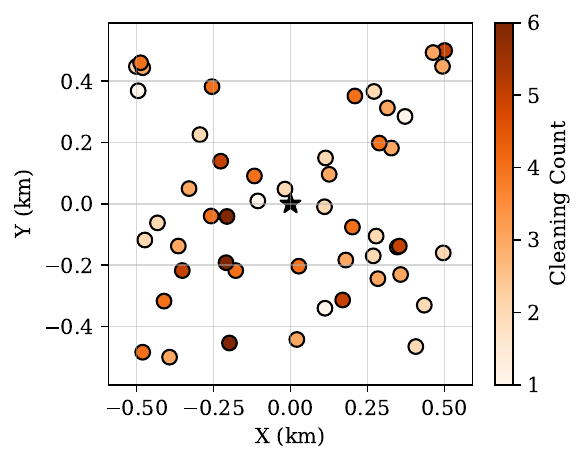}
        \caption{Cleaning activity map}
        \label{fig:cleaning_activity_map}
    \end{subfigure}
    \begin{subfigure}[b]{0.325\textwidth}
        \includegraphics[width=\textwidth]{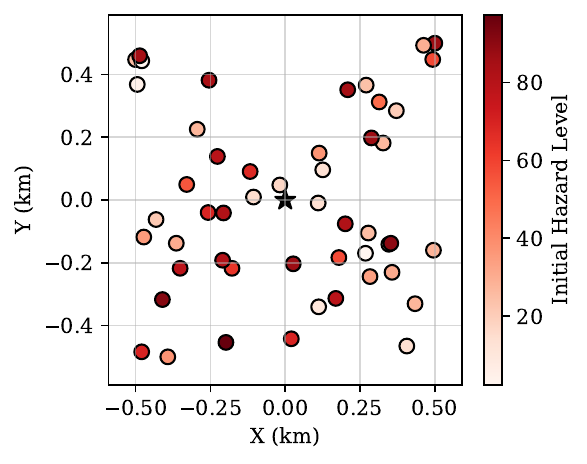}
        \caption{\changed{Initial true hazard map}}
        \label{fig:initial_hazard_map}
    \end{subfigure}

    \caption{Spatial patterns of sensing and cleaning \changed{and initial hazard levels. Site color shows visit frequency or initial hazard, with the depot marked by a star.}}
    \label{fig:activity_maps}
\end{figure}

To illustrate how spatial structure influences task assignment and vehicle routing, we visualize the distribution of sensing and cleaning activities across the environment in \Cref{fig:activity_maps}. These plots show the relationship between site locations, task prioritization, and actual visit frequency throughout the simulation. Once a site was fully cleaned, it was excluded from both sensing and cleaning operations. \changed{Considering these factors, the final activity maps show that sensing visits concentrate on spatially uncertain or high-risk regions, while cleaning routes are primarily allocated to sites with sustained hazard levels until mitigation is complete. This spatial differentiation reduces redundant visits to low-impact sites and allows vehicles to operate near capacity for a larger fraction of the mission, indicating efficient and adaptive task execution throughout the simulation.}

\Cref{fig:sensing_activity_map} reveals a clear spatial pattern. Nearby sites tend to be grouped together in sensing routes, resulting in strong local clustering of visit frequencies. While distance-aware visit value was applied in sensing VRPP, sites closer to the depot were still visited more frequently on average. This is partly because UAVs returning from distant areas could conveniently revisit nearby sites with minimal detour. Conversely, \Cref{fig:cleaning_activity_map} shows a more even distribution of cleaning visits. The number of cleaning visits per site primarily reflects the initial hazard levels, with higher-hazard sites being visited more frequently, \changed{consistent with the spatial distribution of initial true hazards shown in \Cref{fig:initial_hazard_map}.}


\begin{figure}[htbp]
    \centering
    \includegraphics[width=0.5\linewidth]{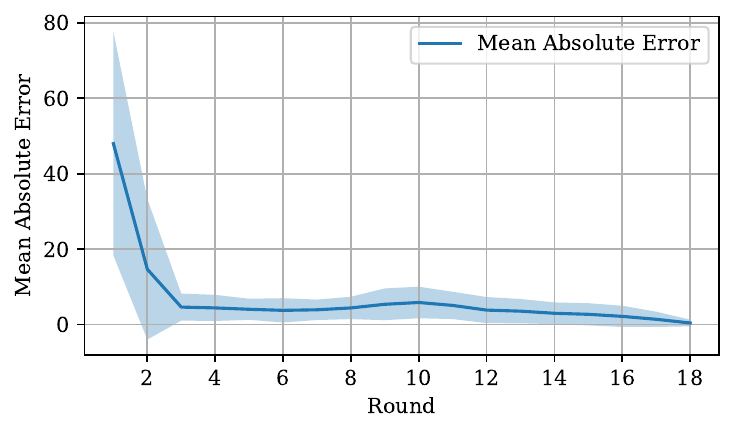}
    \caption{\changed{MAE between estimated and true hazard levels over time. Shaded regions represent the standard deviation across sites.}}
    \label{fig:mae_single}
\end{figure}

\Cref{fig:mae_single} shows the evolution of MAE over time for this single case. The MAE drops sharply during the first three rounds, reflecting the initial exploration phase where many sites are sensed for the first time. After round 3, the error stabilizes at a relatively low level, indicating that the belief model maintains reasonable accuracy. Toward the final rounds, the MAE decreases further as remaining hazards are reduced and beliefs become more certain.

\begin{figure}[htbp]
    \centering

    \begin{subfigure}[b]{0.45\textwidth}
        \includegraphics[width=\textwidth]{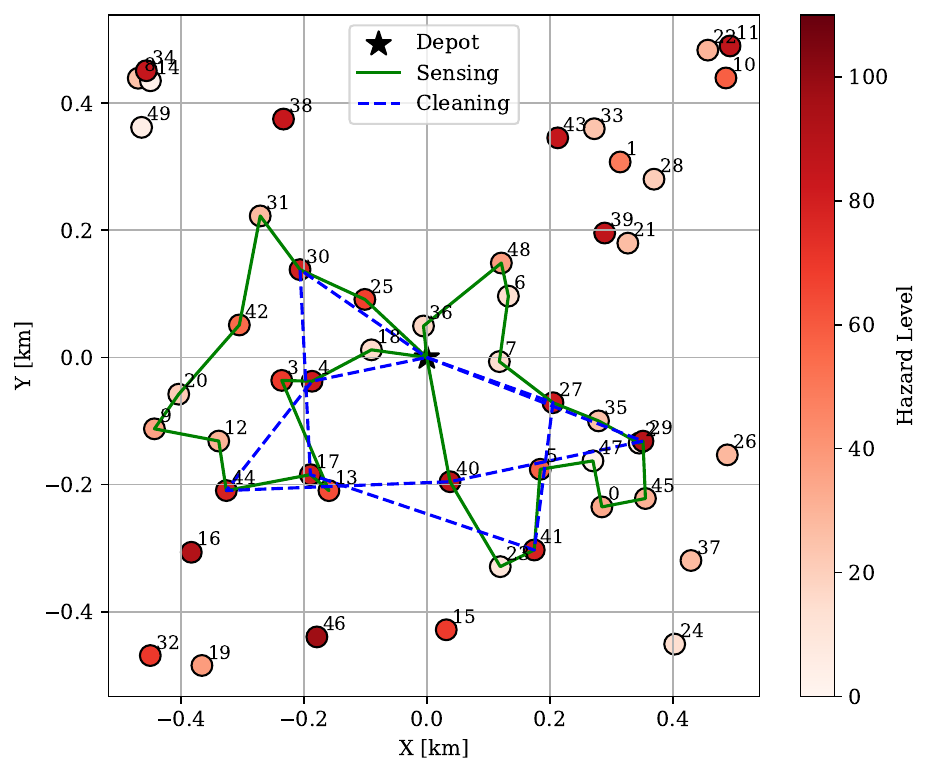}
        \caption{Round 1}
        \label{fig:round1}
    \end{subfigure}
    \hspace{1em}
    \begin{subfigure}[b]{0.45\textwidth}
        \includegraphics[width=\textwidth]{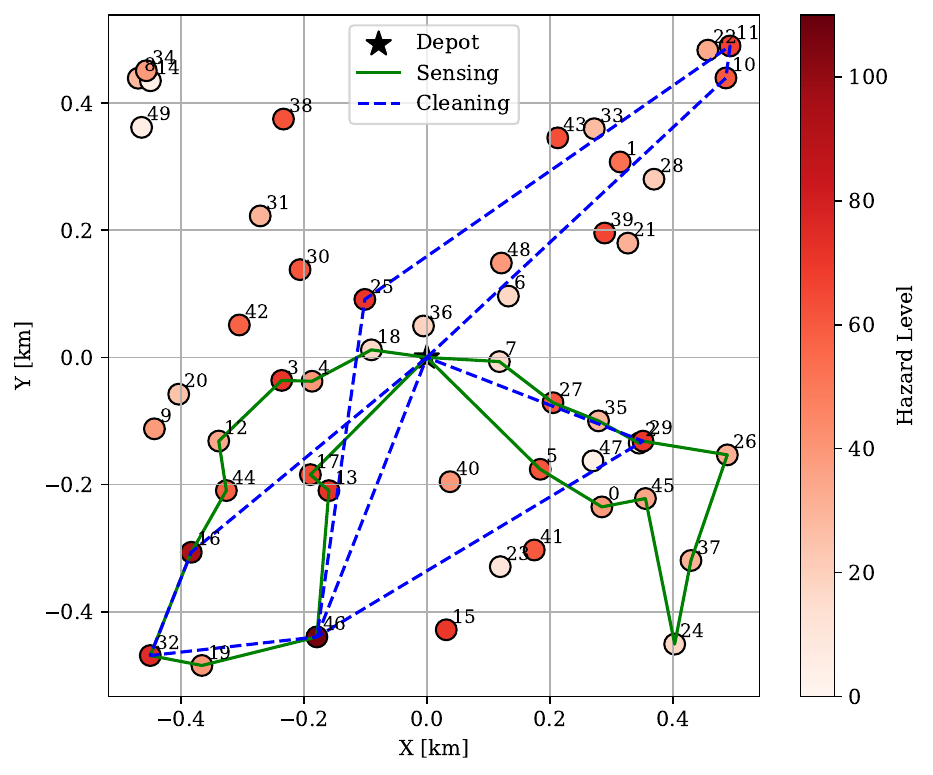}
        \caption{Round 3}
        \label{fig:round3}
    \end{subfigure}

    \vspace{1em}

    \begin{subfigure}[b]{0.45\textwidth}
        \includegraphics[width=\textwidth]{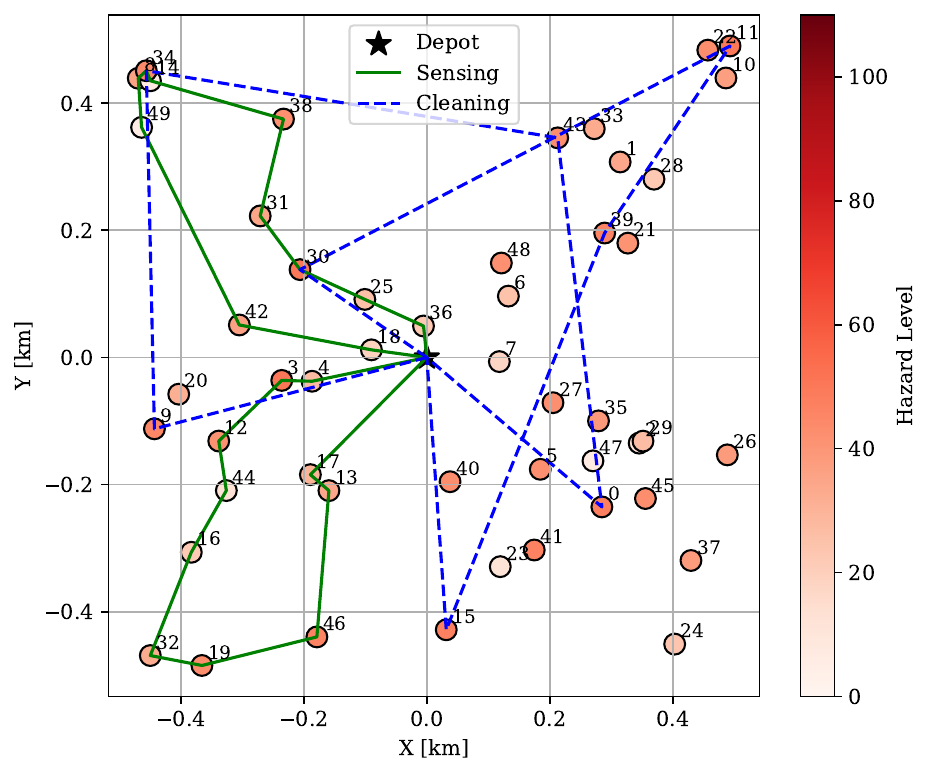}
        \caption{Round 7}
        \label{fig:round7}
    \end{subfigure}
    \hspace{1em}
    \begin{subfigure}[b]{0.45\textwidth}
        \includegraphics[width=\textwidth]{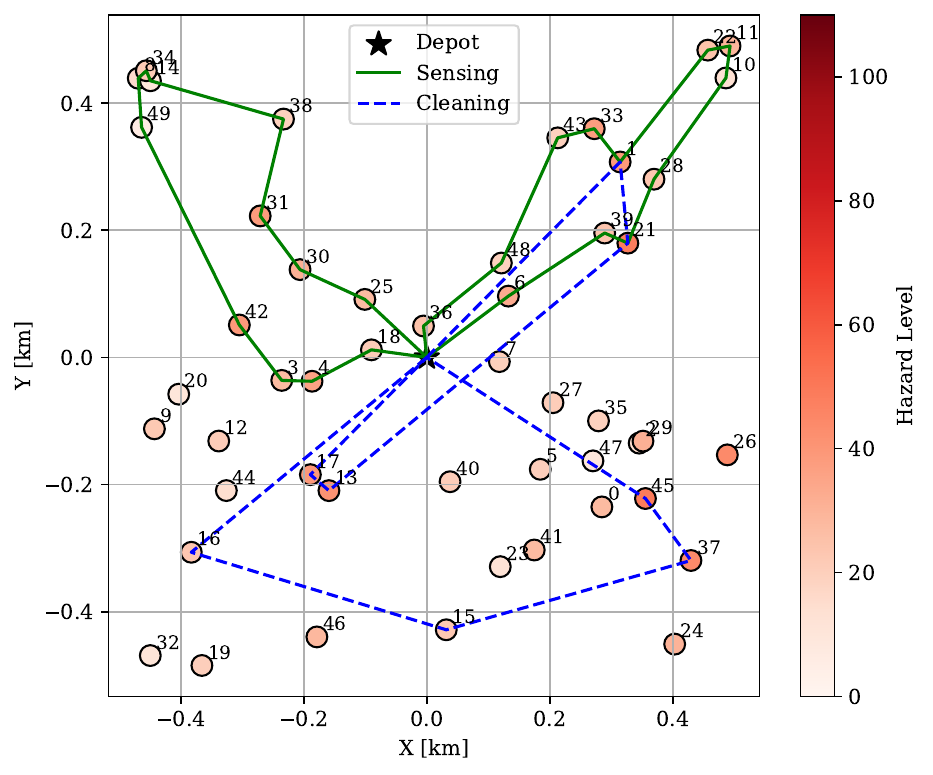}
        \caption{Round 10}
        \label{fig:round10}
    \end{subfigure}

    \vspace{1em}

    \begin{subfigure}[b]{0.45\textwidth}
        \includegraphics[width=\textwidth]{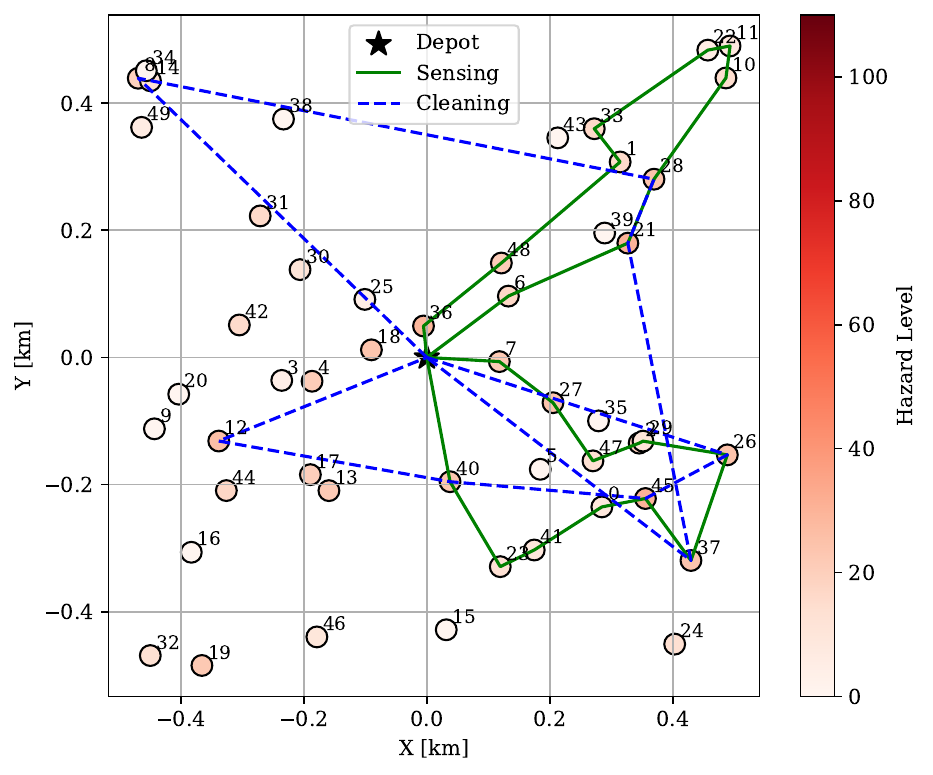}
        \caption{Round 14}
        \label{fig:round14}
    \end{subfigure}
    \hspace{1em}
    \begin{subfigure}[b]{0.45\textwidth}
        \includegraphics[width=\textwidth]{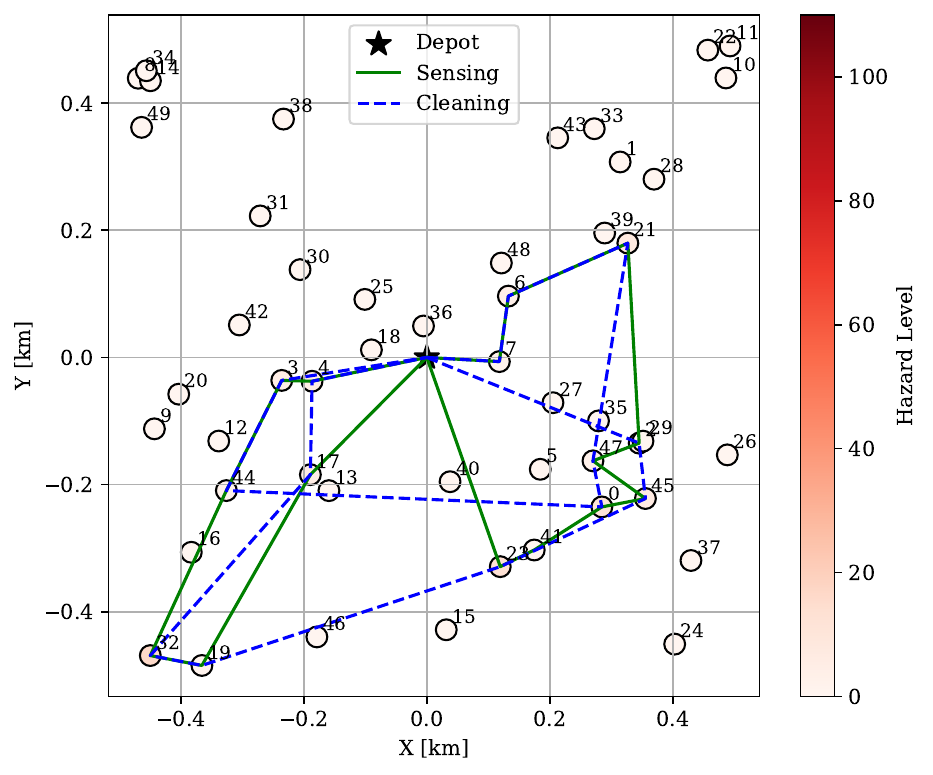}
        \caption{Round 18}
        \label{fig:round18}
    \end{subfigure}
    \caption{
    \changed{Spatiotemporal evolution of UAV and UGV routes. Color intensity shows hazard levels, and lines indicate sensing and cleaning. Animation at \href{https://youtu.be/YCSzIp7RcVA}{youtu.be/YCSzIp7RcVA}.}
    }
    \label{fig:route_evolution}
\end{figure}

To illustrate our framework's operational behavior over time, \Cref{fig:route_evolution} presents selected snapshots. We visualize rounds 1, 3, 7, 10, 14, and 18 to highlight key phases of the decision-making and routing process, achieving complete hazard mitigation by round 18. Across the sequence, the hazard colorbar's maximum value gradually decreases, reflecting successful mitigation as the rounds progress. One exception is observed between \Cref{fig:round1,fig:round3}, where the maximum hazard temporarily increases. As discussed in \Cref{fig:single_reduction}, this is due to the fact that initial rounds did not cover all sites with sensing, allowing a high-hazard site to go undetected. Once identified, it is promptly addressed in the following cleaning phase.

\Cref{fig:round1} shows the system operates in a pure exploration mode. With no prior hazard estimates, UAVs prioritize routes that maximize information gain by visiting spatially clustered sites. UGVs similarly visit only those sites that have been sensed. \Cref{fig:round3} shows a transition to more targeted mitigation. Site 46, the most hazardous, is immediately serviced by both UGVs, enabled by the relaxed cleaning VRPP constraint that allows multiple vehicles per site in a single round. This ensures high-risk sites are promptly addressed.

\Cref{fig:round7,fig:round10,fig:round14} represent the stable operational phase. Hazard levels decline steadily, and BUCB scores guide sensing toward areas needing more information to refine hazard estimates while UGVs operate near full capacity. Routing patterns in these rounds remain spatially efficient, suggesting that distance costs are effectively reflected in the visit value structure of both sensing and cleaning VRPP formulations. \Cref{fig:round18} depicts the final round of the simulation. At this point, all site hazards have dropped below the unit cleaning threshold \( Q_{\text{unit}} \), and UGVs visit many sites with small residual hazards to complete final mitigation. 

\subsubsection{\changed{Sensitivity to Hazard Growth Rate}}

\changed{In the proposed hazard dynamics model, the intrinsic growth rate at each site is sampled as \( \rho_i \sim \mathcal{U}(0,\rho_{\max}) \), introducing heterogeneity in local hazard escalation across the environment. The baseline experiments use \( \rho_{\max} = 0.10 \), as summarized in \Cref{tab:sim_env_params}. To assess robustness with respect to the hazard growth rate, we perform a sensitivity analysis by varying \( \rho_{\max} \in \{0.00, 0.05, 0.10, 0.125, 0.15\} \), while keeping all other model parameters and operational settings fixed. Increasing \( \rho_{\max} \) induces faster hazard escalation and accelerates uncertainty accumulation between sensing rounds, thereby providing a controlled stress test for the uncertainty-aware sensing and routing strategy.}

\begin{table}[htbp]
\centering
\caption{\changed{Sensitivity of termination rounds to hazard growth rate. Termination rounds are shown as median (mean ± std).}}
\label{tab:termination_sensitivity}
\begin{tabular}{lccccc}
\toprule
\textbf{Scenario} &
$\boldsymbol{\rho_{\max}=0.00}$ &
$\boldsymbol{0.05}$ &
$\boldsymbol{0.10}$ &
$\boldsymbol{0.125}$ &
$\boldsymbol{0.15}$ \\
\midrule
Scenario 1 &
7 (7.03 $\pm$ 0.85) &
7 (7.16 $\pm$ 0.90) &
7 (6.82 $\pm$ 1.03) &
8 (7.66 $\pm$ 1.03) &
8 (7.73 $\pm$ 1.06) \\
Scenario 2 &
16 (16.16 $\pm$ 1.13) &
18 (18.17 $\pm$ 1.41) &
20 (20.02 $\pm$ 2.33) &
24 (23.95 $\pm$ 3.14) &
28 (28.33 $\pm$ 5.20) \\
Scenario 3 &
11 (11.28 $\pm$ 0.82) &
12 (12.16 $\pm$ 1.13) &
12 (12.07 $\pm$ 0.99) &
13 (13.47 $\pm$ 1.20) &
14 (14.35 $\pm$ 1.43) \\
\bottomrule
\end{tabular}
\end{table}

\changed{As shown in \Cref{tab:termination_sensitivity}, increasing the hazard growth rate generally leads to longer termination rounds across all scenarios, since hazards escalate more rapidly between sensing rounds. Despite this effect, the proposed framework consistently converges and achieves full hazard clearance for all tested growth-rate settings, demonstrating robust behavior. The impact of growth-rate changes differs by scenario. In the smallest environment (Scenario~1), termination rounds change only marginally as the growth rate increases, indicating that system performance is dominated by early saturation effects. In larger environments (Scenarios~2 and~3), termination rounds increase more noticeably, as faster hazard growth leads to greater uncertainty over larger spatial areas. Nevertheless, performance trends remain stable and predictable across all scenarios.}

\subsection{Comparison Against Baselines}
\subsubsection{Baseline Algorithms}
To evaluate the effectiveness of the proposed BUCB-based sensing strategy, we compare it against \changed{four} baseline methods that differ in the way sensing scores are assigned. All other components are kept consistent across experiments to ensure a fair comparison. In particular, the underlying VRPP formulation, including constraints for both sensing and cleaning operations, is shared across all strategies. Cleaning decisions are based on the estimated hazard level \( \mu_i \) at each site. While BUCB additionally incorporates the uncertainty term \( \sigma_i^2 \) in both sensing and cleaning decisions, the baseline methods rely solely on \( \mu_i \) and do not model uncertainty. The \changed{four} baseline strategies are described below.

\begin{itemize}
  \item \changed{\text{Point-Estimate Strategy:}
  This baseline assigns sensing scores using only point estimates of site hazards. Each site $i$ is associated with a single estimated hazard value $\mu_i$, corresponding to the most recent sensing observation. Unvisited sites are prioritized to encourage initial exploration, and the sensing VRPP is solved using these point estimates as visit values. No uncertainty or variance information is incorporated in the decision-making process.}
  
  \item \text{Random Strategy:}  
  At each round, a random score is assigned to every site with a nonzero hazard. The score is drawn independently from a uniform distribution \(\mathcal{U}(0,1)\) and used as the visit value in the sensing VRPP. This strategy represents an uninformed policy that makes no use of past observations or hazard estimates.

  \item \text{Round-Robin Strategy:}  
  This strategy prioritizes sites that have been selected less frequently in previous rounds. Each site is scored inversely proportional to its cumulative sensing count, relative to the site that is visited most frequently. The resulting score is used as the sensing visit value in the VRPP. Although it promotes spatial fairness, the method does not account for actual hazard levels or uncertainty.

  \item \text{Oracle Strategy:}  
  This baseline assumes access to the true hazard values at all sites. At every round, the belief mean \(\mu_i\) is set equal to the ground-truth hazard. This removes all sensing noise and eliminates uncertainty. While not feasible in practice, the oracle serves as an upper-bound reference under perfect information.
\end{itemize}

\subsubsection{Quantitative Performance Comparison}

We evaluate all strategies under the same three simulation settings introduced in \Cref{sec:proposed_method}, which vary in the number of sites and the number of available sensing and cleaning drones. For each scenario, we run 100 trials using different random seeds. The initial site layout, hazard dynamics, and VRP constraints are fixed for each simulation to ensure consistent evaluation across strategies. \Cref{tab:summary_metrics} summarizes the performance metrics across all strategies and scenarios.

\begin{table}[htbp]
\centering
\caption{Performance comparison across strategies and scenarios. \changed{Oracle denotes an upper bound; bold values indicate the closest strategy, termination rounds shown as median (mean ± std).}}
\begin{tabular}{c l ccc}
\toprule
\textbf{Scenario} & \textbf{Strategy} & \textbf{Termination Rounds} & \textbf{Cumulative Hazard} & \textbf{Cleaning Rate} \\
\midrule
\multirow{5}{*}{Scenario 1} 
  & \textit{Oracle}      & \textit{6  (5.87 ± 0.92)}  & \textit{3280.98 ± 930.16} & \textit{162.81 ± 13.34} \\
  & BUCB        & \textbf{7  (6.82 ± 1.03)}  & \textbf{3355.02 ± 933.25} & \textbf{160.94 ± 14.63} \\
  & \changed{Point-Estimate} 
    & \changed{8 (7.58 ± 1.57)} 
    & \changed{3418.55 ± 980.38} 
    & \changed{142.79 ± 17.55} \\ 
  & Random      & 10 (10.45 ± 1.98) & 3427.64 ± 953.87 &  92.07 ± 19.61 \\ 
  & Round-Robin & 11 (11.16 ± 2.00) & 3377.70 ± 919.78 &  91.17 ± 20.63 \\
\midrule
\multirow{4}{*}{Scenario 2} 
  & \textit{Oracle}      & \textit{18 (17.86 ± 2.11)} & \textit{25364.91 ± 5020.53} & \textit{187.29 ±  6.15} \\
  & BUCB        & \textbf{20 (20.02 ± 2.33)} & 26150.08 ± 5186.27 & \textbf{187.73 ±  7.53} \\
  & \changed{Point-Estimate} 
    & \changed{24 (23.62 ± 3.36)} 
    & \changed{25726.01 ± 5865.95} 
    & \changed{165.06 ± 16.19}\\ 
  & Random      & 24 (23.98 ± 2.75) & \textbf{25632.98 ± 4803.43} & 134.55 ± 12.43 \\
  & Round-Robin & 26 (26.15 ± 3.22) & 25787.71 ± 4930.72 & 135.11 ± 13.36 \\
\midrule
\multirow{4}{*}{Scenario 3} 
  & \textit{Oracle}      & \textit{10 (10.45 ± 1.03)} & \textit{14491.85 ± 2412.11} & \textit{265.68 ± 10.01} \\
  & BUCB        & \textbf{12 (12.07 ± 0.99)} & 14863.46 ± 2420.60 & \textbf{260.52 ± 13.10} \\
  & \changed{Point-Estimate} 
    & \changed{15 (14.70 ± 2.05)} 
    & \changed{149518.12 ± 2411.30} 
    & \changed{209.01 ± 16.61} \\
  & Random      & 18 (17.98 ± 2.85) & 15039.73 ± 2374.54 & 142.61 ± 24.50 \\
  & Round-Robin & 19 (18.91 ± 2.77) & \textbf{14859.73 ± 2429.83} & 153.71 ± 20.99 \\
\bottomrule
\end{tabular}
\label{tab:summary_metrics}
\end{table}

Across all scenarios, BUCB consistently demonstrates strong performance, almost matching that of the Oracle strategy. In terms of termination rounds, BUCB requires only one or two more rounds than the Oracle across all scenarios, and significantly fewer than the Random and Round-Robin strategies. \changed{The Point-Estimate strategy performs better than the uninformed baselines, but still requires more rounds than BUCB, particularly in larger scenarios. This intermediate performance reflects its reliance on deterministic hazard estimates without explicit exploration.} For example, in Scenario 1, BUCB completes the task in a median of 7 rounds compared to \changed{8,} 10 and 11 rounds for the \changed{Point-Estimate,} Random and Round-Robin strategies, respectively. In addition, BUCB achieves a substantially higher cleaning rate than the two uninformed baselines \changed{and also outperforms the Point-Estimate strategy, highlighting the benefit of incorporating uncertainty into sensing decisions.} In Scenario 3, for instance, BUCB maintains an average cleaning rate of 260.5, compared to \changed{210.5,} 142.6 and 153.7 for the \changed{Point-Estimate,} Random and Round-Robin strategies, respectively. These results indicate that uncertainty-aware sensing enables more efficient allocation of cleaning resources, particularly under more demanding conditions. These findings demonstrate that BUCB effectively balances exploration and exploitation, leading to faster hazard elimination and more efficient system operation compared to strategies that do not model uncertainty.

However, no single strategy consistently achieves cumulative hazard values closest to the Oracle across all scenarios. BUCB shows hazard values close to Oracle in Scenario 1, but in Scenarios 2 and 3, Random and Round-Robin are more comparable, indicating no consistent advantage for BUCB. This suggests that while BUCB is generally more efficient in reaching task completion, its advantage in cumulative hazard varies depending on the scenario. One likely reason is that all strategies share the same VRPP formulation, which may reduce the impact of differences in sensing decisions. Since the fleet configurations are also identical, the cleaning efficiency is effectively held constant. In this context, BUCB’s tendency to explore more in the early stages to support more efficient cleaning later may result in greater hazard accumulation early on. This limits its advantage in cumulative hazard, even though it completes the overall task more quickly.



\begin{figure}[htbp]
    \centering

    \begin{subfigure}[b]{0.48\linewidth}
        \centering
        \includegraphics[width=\linewidth]{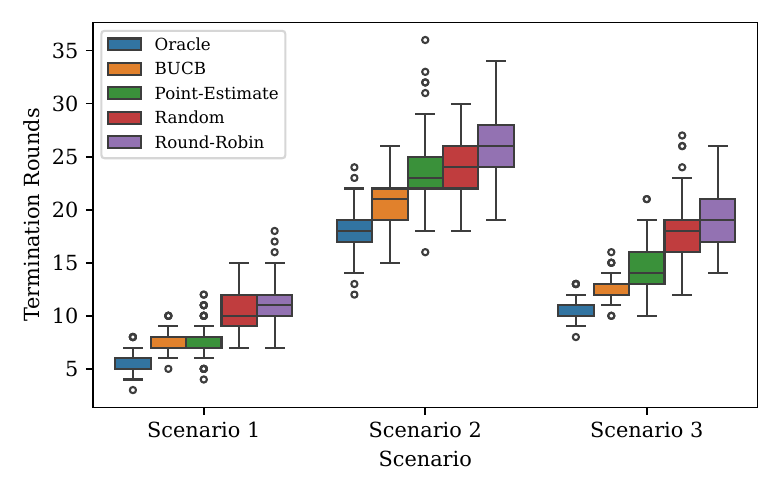}
        \caption{\changed{Termination rounds across strategies and scenarios.}}
        \label{fig:termination_rounds_boxplot}
    \end{subfigure}
    \hfill
    \begin{subfigure}[b]{0.48\linewidth}
        \centering
        \includegraphics[width=\linewidth]{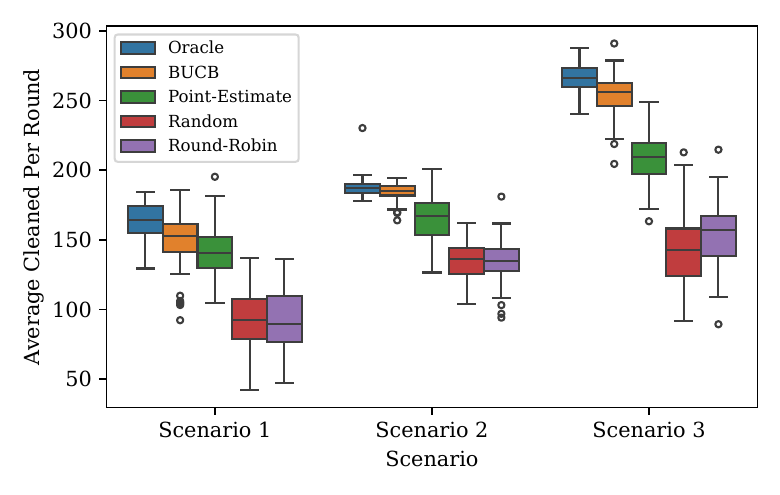}
        \caption{\changed{Average cleaning rate per round by strategy.}}
        \label{fig:cleaning_rate_boxplot}
    \end{subfigure}

    \caption{\changed{
    Performance comparison across strategies. \changed{BUCB completes tasks in fewer rounds than baselines, approaching Oracle performance with higher average cleaning rates.}}
    }
    \label{fig:strategy_performance_boxplots}
\end{figure}

\Cref{fig:termination_rounds_boxplot,fig:cleaning_rate_boxplot} visualize the distribution of results across trials, clarifying the stability of each strategy’s performance. BUCB demonstrates consistently low variability in termination rounds across all scenarios, indicating stable task completion performance. In contrast, the Random and Round-Robin strategies exhibit higher variance, suggesting more inconsistent outcomes across trials. \changed{The Point-Estimate strategy exhibits moderate variability, remaining more stable than the uninformed baselines but less consistent than BUCB.} A similar trend is observed in cleaning performance. As shown in \Cref{fig:cleaning_rate_boxplot}, BUCB achieves a high average cleaning rate with relatively small standard deviations, while Random and Round-Robin exhibit significantly greater spread. These results reinforce BUCB’s reliability in achieving consistent resource allocation and cleaning efficiency.

\section{Conclusion and Future Work}
\subsection{Concluding Remarks} \label{sec:conclusion}

We present a decision-making framework that combines a BUCB-based exploration-exploitation strategy with vehicle routing to support adaptive UAV/UGV dispatch in hazardous environments. The system maintains site-level hazard estimates using time-weighted Bayesian updates and extrapolates future risk based on recent trends. This allows it to plan under uncertainty while focusing attention on high-risk or poorly observed areas. Sensing and cleaning phases are handled through separate VRPP formulations, each reflecting task-specific priorities. The routing strategy accounts for site-level constraints and vehicle capacities, enabling practical operation over large environments.

The simulations demonstrate that the proposed method successfully eliminates all hazards, simultaneously achieving confident estimations and efficient vehicle use. These results hold across different scales and fleet settings. The findings highlight how uncertainty-aware scoring and constrained routing can work together to effectively manage dynamic hazards. Compared to uninformed baselines such as Random or Round-Robin sensing, the proposed approach completes the task significantly faster and with higher cleaning efficiency, while maintaining low trial-to-trial variability. This indicates that explicitly modeling uncertainty not only improves average performance but also enhances reliability under varying conditions. \changed{This advantage is further supported by comparisons with a Point-Estimate baseline, where the absence of uncertainty-aware exploration leads to slower convergence and greater variability. Taken together, these results suggest that the proposed framework achieves robust and efficient performance across the tested environmental settings and vehicle capabilities.} The proposed framework offers a flexible foundation for dynamic hazard monitoring and mitigation. The framework can be easily generalized to other multi-agent planning problems that involve uncertainty, limited resources, and evolving environmental states.

\subsection{Limitation of Work and Future Directions} 
\label{sec:future_work}
While the proposed framework demonstrates effective performance in simulated scenarios, several simplifying assumptions limit its generalizability and real-world applicability. This section outlines key limitations and describes possible directions for future research.

First, the model assumes that hazard levels evolve according to a spatially coupled logistic process. While this structure captures basic nonlinear growth and spatial interactions, it cannot represent more complex phenomena such as wind-driven transport, terrain-induced diffusion, or delayed hazard propagation. Future work could incorporate more expressive dynamics using physics-based models or learned hazard evolution processes informed by domain-specific data.

Second, the belief update mechanism relies on gradient-based extrapolation for unobserved sites. Although this approach is simple, efficient, and responsive to local trends, it becomes unreliable when observations are sparse or highly noisy. Enhancing belief propagation using statistical filters, structured priors, or learning-based predictors could improve robustness and estimation accuracy in partially observed environments.

Third, the current implementation uses a fixed number of UAVs and UGVs per round, with predefined roles for sensing and cleaning. This rigid fleet structure may limit adaptability in missions with varying hazard severity, spatial extent, or vehicle availability. A promising extension is to enable dynamic fleet assignment, allowing the system to allocate resources across tasks based on real-time operational conditions flexibly. \changed{Explicit travel-time or service-time constraints can be handled through a VRPTW formulation by substituting the routing solver within the proposed framework.}

Additionally, the current framework separates sensing and cleaning into distinct operations. A natural extension would be to allow simultaneous sensing and cleaning, which could shorten mission timelines and reduce redundant site visits, though it would also introduce added computational and coordination complexity. Another related direction is to impose a limit on the number of cleaning rounds per site, pushing the system to prioritize more accurate sensing upfront and motivating the development of hybrid UGVs capable of lightweight sensing during cleaning to verify hazard levels in real time.

Finally, the framework assumes centralized coordination with full observability and instantaneous communication. Real-world deployments often involve limited bandwidth, latency, or partial information sharing between agents. Future work could explore decentralized or hierarchical planning strategies where agents operate with delayed or local information. Communication-aware coordination would further improve robustness in distributed and constrained environments.

\section*{Acknowledgments}
This project was supported by NSF IUCRC Phase I: Center for Autonomous Air Mobility and Sensing (CAAMS) Award No. 2137195. The authors would like to thank Isaac Weintraub, David Casbeer, and Alexander von Moll from the Air Force Research Laboratory (AFRL), as well as Begum Cannataro from Draper, for providing valuable feedback and comments from the early stages of this work. We also thank Cameron K. Peterson and Grant Stagg for helpful discussions.

\bibliography{sample}

@inproceedings{le_ny_multi-uav_2008,
	address = {Seattle, WA},
	title = {Multi-{UAV} dynamic routing with partial observations using restless bandit allocation indices},
	isbn = {978-1-4244-2078-0},
	doi = {10.1109/ACC.2008.4587156},
	language = {en},
	urldate = {2024-10-17},
	booktitle = {2008 {American} {Control} {Conference}},
	publisher = {IEEE},
	author = {Le Ny, Jerome and Dahleh, Munther and Feron, Eric},
	month = jun,
	year = {2008},
	pages = {4220--4225},
}

@article{lagos_multi-armed_2024,
	title = {Multi-armed bandit-based hyper-heuristics for combinatorial optimization problems},
	volume = {312},
	issn = {03772217},
	doi = {10.1016/j.ejor.2023.06.016},
	language = {en},
	number = {1},
	urldate = {2024-10-17},
	journal = {European Journal of Operational Research},
	author = {Lagos, Felipe and Pereira, Jordi},
	month = jan,
	year = {2024},
	pages = {70--91},
}

@article{harrath_algorithm_2024,
	title = {An {Algorithm} {Based} on {Priority} {Rules} for {Solving} a {Multi}-drone {Routing} {Problem} in {Hazardous} {Waste} {Collection}},
	volume = {15},
	issn = {21565570, 2158107X},
	doi = {10.14569/IJACSA.2024.0150298},
	language = {en},
	number = {2},
	urldate = {2024-09-13},
	journal = {International Journal of Advanced Computer Science and Applications},
	author = {Harrath, Youssef and Kaabi, Jihene and Alaradi, Eman and Alnoaimi, Manar and Alawadhi, Noor},
	year = {2024},
}

@article{abdulsattar_ant_2023,
	title = {Ant colony algorithm to solve a drone routing problem for hazardous waste collection},
	volume = {30},
	issn = {2576-5299},
	doi = {10.1080/25765299.2023.2275505},
	language = {en},
	number = {1},
	urldate = {2024-09-13},
	journal = {Arab Journal of Basic and Applied Sciences},
	author = {Abdulsattar, Khadija and Harrath, Youssef and Kaabi, Jihene},
	month = dec,
	year = {2023},
	pages = {636--649},
}

@article{kaabi_2-phase_2023,
	title = {A 2-phase approach for planning of hazardous waste collection using an unmanned aerial vehicle},
	volume = {21},
	issn = {1619-4500, 1614-2411},
	doi = {10.1007/s10288-022-00526-0},
	language = {en},
	number = {4},
	urldate = {2024-09-13},
	journal = {4OR},
	author = {Kaabi, Jihene and Harrath, Youssef and Mahjoub, Amine and Hewahi, Nabil and Abdulsattar, Khadija},
	month = dec,
	year = {2023},
	pages = {585--608},
}

@article{kas_using_2020,
	title = {Using unmanned aerial vehicles and robotics in hazardous locations safely},
	volume = {39},
	issn = {1066-8527, 1547-5913},
	doi = {10.1002/prs.12066},
	language = {en},
	number = {1},
	urldate = {2024-09-13},
	journal = {Process Safety Progress},
	author = {Kas, Kathleen A. and Johnson, Gary K.},
	month = mar,
	year = {2020},
	pages = {e12066},
}

@article{liang_waste_2022,
	title = {Waste collection routing problem: {A} mini-review of recent heuristic approaches and applications},
	volume = {40},
	issn = {0734-242X, 1096-3669},
	shorttitle = {Waste collection routing problem},
	doi = {10.1177/0734242X211003975},
	language = {en},
	number = {5},
	urldate = {2024-10-20},
	journal = {Waste Management \& Research: The Journal for a Sustainable Circular Economy},
	author = {Liang, Yun-Chia and Minanda, Vanny and Gunawan, Aldy},
	month = may,
	year = {2022},
	pages = {519--537},
}

@article{dotoli_vehicle_2017,
	title = {A {Vehicle} {Routing} {Technique} for {Hazardous} {Waste} {Collection}},
	volume = {50},
	copyright = {https://www.elsevier.com/tdm/userlicense/1.0/},
	issn = {24058963},
	doi = {10.1016/j.ifacol.2017.08.2051},
	language = {en},
	number = {1},
	urldate = {2024-09-13},
	journal = {IFAC-PapersOnLine},
	author = {Dotoli, Mariagrazia and Epicoco, Nicola},
	month = jul,
	year = {2017},
	pages = {9694--9699},
}

@article{suksee_grasp_2021,
	title = {{GRASP} with {ALNS} for solving the location routing problem of infectious waste collection in the {Northeast} of {Thailand}},
	volume = {12},
	issn = {19232926, 19232934},
	doi = {10.5267/j.ijiec.2021.2.001},
	language = {en},
	number = {3},
	urldate = {2024-09-13},
	journal = {International Journal of Industrial Engineering Computations},
	author = {Suksee, Siwaporn and Sindhuchao, Sombat},
	year = {2021},
	pages = {305--320},
}

@article{jaselskis_robotic_1994,
	title = {Robotic {Opportunities} for {Hazardous}‐{Waste} {Cleanup}},
	volume = {120},
	issn = {0733-9372, 1943-7870},
	doi = {10.1061/(ASCE)0733-9372(1994)120:2(359)},
	language = {en},
	number = {2},
	urldate = {2024-09-13},
	journal = {Journal of Environmental Engineering},
	author = {Jaselskis, Edward J. and Anderson, Mary Rose},
	month = mar,
	year = {1994},
	pages = {359--378},
}

@article{rabbani_using_2021,
	title = {Using modified metaheuristic algorithms to solve a hazardous waste collection problem considering workload balancing and service time windows},
	volume = {25},
	issn = {1432-7643, 1433-7479},
	doi = {10.1007/s00500-020-05261-4},
	language = {en},
	number = {3},
	urldate = {2024-09-13},
	journal = {Soft Computing},
	author = {Rabbani, Masoud and Nikoubin, Alireza and Farrokhi-Asl, Hamed},
	month = feb,
	year = {2021},
	pages = {1885--1912},
}

@article{kim_waste_2006,
	title = {Waste collection vehicle routing problem with time windows},
	volume = {33},
	copyright = {https://www.elsevier.com/tdm/userlicense/1.0/},
	issn = {03050548},
	doi = {10.1016/j.cor.2005.02.045},
	language = {en},
	number = {12},
	urldate = {2024-09-13},
	journal = {Computers \& Operations Research},
	author = {Kim, Byung-In and Kim, Seongbae and Sahoo, Surya},
	month = dec,
	year = {2006},
	pages = {3624--3642},
}

@article{shakhatreh2019unmanned,
  title={Unmanned aerial vehicles (UAVs): A survey on civil applications and key research challenges},
  author={Shakhatreh, Hazim and Sawalmeh, Ahmad H and Al-Fuqaha, Ala and Dou, Zuochao and Almaita, Eyad and Khalil, Issa and Othman, Noor Shamsiah and Khreishah, Abdallah and Guizani, Mohsen},
  journal={IEEE Access},
  doi={10.1109/ACCESS.2019.2909530},
  volume={7},
  pages={48572--48634},
  year={2019},
  publisher={IEEE}
}

@inproceedings{choi_adaptive_rmdp_2026,
author = {Jimin Choi and Mengmeng Li and Max Z. Li},
title = {Adaptive Robust Markov Decision Process for Wide-Area Surveillance With Collaborative Combat Aircraft},
booktitle = {AIAA SCITECH 2026 Forum},
  year      = {2026},
  month     = {January},
doi = {10.2514/6.2026-2884},
}

@INPROCEEDINGS{bilevel2025,
  author={Choi, Jimin and Stagg, Grant and Peterson, Cameron K. and Li, Max Z.},
  booktitle={2025 IEEE 64th Conference on Decision and Control (CDC)}, 
  title={Bi-Level Route Optimization and Path Planning with Hazard Exploration}, 
  year={2025},
  volume={},
  number={},
  pages={3403-3410},
  doi={10.1109/CDC57313.2025.11312588}}

@InProceedings{pmlr-v22-kaufmann12,
  title = 	 {On Bayesian Upper Confidence Bounds for Bandit Problems},
  author = 	 {Kaufmann, Emilie and Cappe, Olivier and Garivier, Aurelien},
  booktitle = 	 {Proceedings of the Fifteenth International Conference on Artificial Intelligence and Statistics},
  pages = 	 {592--600},
  year = 	 {2012},
  editor = 	 {Lawrence, Neil D. and Girolami, Mark},
  volume = 	 {22},
  series = 	 {Proceedings of Machine Learning Research},
  address = 	 {La Palma, Canary Islands},
  month = 	 {21--23 Apr},
  publisher =    {PMLR},
}

@book{gelman2013bayesian,
  title        = {Bayesian Data Analysis},
  author       = {Gelman, Andrew and Carlin, John B. and Stern, Hal S. and Dunson, David B. and Vehtari, Aki and Rubin, Donald B.},
  year         = {2013},
  edition      = {3rd},
  publisher    = {Chapman \& Hall/CRC},
  address      = {Boca Raton, FL},
  isbn         = {978-1439840955},
  pages        = {675},
  language     = {English}
}

@article{pyvrp,
  doi = {10.1287/ijoc.2023.0055},
  year = {2024},
  volume = {36},
  number = {4},
  pages = {943--955},
  publisher = {INFORMS},
  author = {Niels A. Wouda and Leon Lan and Wouter Kool},
  title = {{PyVRP}: a high-performance {VRP} solver package},
  journal = {INFORMS Journal on Computing},
}

@INPROCEEDINGS{8877114,
  author={La Scalea, R. and Rodrigues, M. and Osorio, D. P. M. and Lima, C. H. and Souza, R. D. and Alves, H. and Branco, K. C.},
  booktitle={2019 16th International Symposium on Wireless Communication Systems (ISWCS)}, 
  title={Opportunities for autonomous UAV in harsh environments}, 
  year={2019},
  volume={},
  number={},
  pages={227-232},
  doi={10.1109/ISWCS.2019.8877114}}

@Article{smartcities8010005,
AUTHOR = {Liang, Jun and Zhang, Zongjia and Zhi, Yanpeng},
TITLE = {Multi-Armed Bandit Approaches for Location Planning with Dynamic Relief Supplies Allocation Under Disaster Uncertainty},
JOURNAL = {Smart Cities},
VOLUME = {8},
YEAR = {2025},
NUMBER = {1},
ARTICLE-NUMBER = {5},
ISSN = {2624-6511},
DOI = {10.3390/smartcities8010005}
}

@ARTICLE{8669870,
  author={Lin, Yu and Wang, Tianyu and Wang, Shaowei},
  journal={IEEE Communications Letters}, 
  title={UAV-Assisted Emergency Communications: An Extended Multi-Armed Bandit Perspective}, 
  year={2019},
  volume={23},
  number={5},
  pages={938-941},
  doi={10.1109/LCOMM.2019.2906194}}

@Article{s23031402,
AUTHOR = {Amrallah, Amr and Mohamed, Ehab Mahmoud and Tran, Gia Khanh and Sakaguchi, Kei},
TITLE = {UAV Trajectory Optimization in a Post-Disaster Area Using Dual Energy-Aware Bandits},
JOURNAL = {Sensors},
VOLUME = {23},
YEAR = {2023},
NUMBER = {3},
ARTICLE-NUMBER = {1402},
PubMedID = {36772443},
ISSN = {1424-8220},
DOI = {10.3390/s23031402}
}

@article{MAL-068,
year = {2019},
volume = {12},
journal = {Foundations and Trends® in Machine Learning},
title = {Introduction to Multi-Armed Bandits},
doi = {10.1561/2200000068},
issn = {1935-8237},
number = {1-2},
pages = {1-286},
author = {Aleksandrs Slivkins }
}

@misc{dji_vietnam_2021,
  author       = {{DJI Enterprise}},
  title        = {Drone Search and Rescue in Vietnam: How Drones Helped During Floods and Landslides},
  year         = {2021},
  url          = {https://enterprise-insights.dji.com/user-stories/drone-search-and-rescue-vietnam-flood-landslides},
  note         = {Accessed: 2025-07-14}
}

@misc{terra_drone_angola_2019,
  author       = {{sUAS News}},
  title        = {Terra Drone Angola uses UAV in offshore mock oil spill response},
  year         = {2019},
  url          = {https://www.suasnews.com/2019/09/terra-drone-angola-uses-uav-in-offshore-mock-oil-spill-response/},
  note         = {Accessed: 2025-07-14}
}

@misc{rheinmetall_mission_master,
  author       = {{Rheinmetall}},
  title        = {Mission Master – Uncrewed Ground Vehicles (UGVs)},
  year         = {2024},
  url          = {https://www.rheinmetall.com/en/products/uncrewed-vehicles/uncrewed-ground-systems/mission-master-a-ugs},
  note         = {Accessed: 2025-07-14}
}

@InProceedings{10.1007/978-3-319-12145-1_12,
author="Kozitsky, Yuri",
editor="Banasiak, Jacek
and Bobrowski, Adam
and Lachowicz, Miros{\l}aw",
title="Dynamics of Spatial Logistic Model: Finite Systems",
booktitle="Semigroups of Operators -Theory and Applications",
year="2015",
publisher="Springer International Publishing",
address="Cham",
pages="197--211",
isbn="978-3-319-12145-1",
doi="https://doi.org/10.1007/978-3-319-12145-1_12"
}

@InProceedings{10.1007/3-540-59496-5_337,
author="Jakobi, Nick
and Husbands, Phil
and Harvey, Inman",
editor="Mor{\'a}n, Federico
and Moreno, Alvaro
and Merelo, Juan Juli{\'a}n
and Chac{\'o}n, Pablo",
title="Noise and the reality gap: The use of simulation in evolutionary robotics",
booktitle="Advances in Artificial Life",
year="1995",
publisher="Springer Berlin Heidelberg",
address="Berlin, Heidelberg",
pages="704--720",
isbn="978-3-540-49286-3",
doi="https://doi.org/10.1007/3-540-59496-5_337"
}

@Article{robotics10020078,
AUTHOR = {Groves, Keir and Hernandez, Emili and West, Andrew and Wright, Thomas and Lennox, Barry},
TITLE = {Robotic Exploration of an Unknown Nuclear Environment Using Radiation Informed Autonomous Navigation},
JOURNAL = {Robotics},
VOLUME = {10},
YEAR = {2021},
NUMBER = {2},
ARTICLE-NUMBER = {78},
ISSN = {2218-6581},
DOI = {10.3390/robotics10020078}
}

@misc{schwaiger2024ugvcbrnunmannedgroundvehicle,
      title={UGV-CBRN: An Unmanned Ground Vehicle for Chemical, Biological, Radiological, and Nuclear Disaster Response}, 
      author={Simon Schwaiger and Lucas Muster and Georg Novotny and Michael Schebek and Wilfried Wöber and Stefan Thalhammer and Christoph Böhm},
      year={2024},
      eprint={2406.14385},
      archivePrefix={arXiv},
      primaryClass={cs.RO},
      url={https://arxiv.org/abs/2406.14385}, 
}

@inproceedings{choi2025bandit,
  author    = {Jimin Choi and Max Z. Li},
  title     = {Bandit-Enabled Dynamic Vehicle Dispatch and Routing in Hazardous Environments},
  booktitle = {AIAA AVIATION FORUM AND ASCEND 2025},
  year      = {2025},
  month     = {July},
  doi       = {10.2514/6.2025-3411},
}

@misc{hazwoper_osha_2025,
  title        = {What are the Most Common Accidents in Remediation Work and How to Prevent Them?},
  author       = {{HAZWOPER OSHA Training, LLC}},
  year         = {2025},
  month        = {July},
  url          = {https://hazwoper-osha.com/blog-post/common-accidents-in-remediation-work-and-how-to-prevent-them},
  note         = {Accessed: 2025-08-01}
}

@article{LEE2006265,
title = {A shortest path approach to the multiple-vehicle routing problem with split pick-ups},
journal = {Transportation Research Part B: Methodological},
volume = {40},
number = {4},
pages = {265-284},
year = {2006},
issn = {0191-2615},
doi = {https://doi.org/10.1016/j.trb.2004.11.004},
author = {Chi-Guhn Lee and Marina A. Epelman and Chelsea C. White and Yavuz A. Bozer},
}

@article{0278cc85-5bd0-38d1-b78b-21ed2d63bafc,
 ISSN = {00411655, 15265447},
 author = {C. Archetti and M. G. Speranza and A. Hertz},
 journal = {Transportation Science},
 number = {1},
 pages = {64--73},
 publisher = {INFORMS},
 title = {A Tabu Search Algorithm for the Split Delivery Vehicle Routing Problem},
 urldate = {2025-08-01},
 volume = {40},
 year = {2006},
 doi = {https://doi.org/10.1016/j.cor.2018.07.021}
}

@article{10.1145/321043.321046,
author = {Miller, C. E. and Tucker, A. W. and Zemlin, R. A.},
title = {Integer Programming Formulation of Traveling Salesman Problems},
year = {1960},
issue_date = {Oct. 1960},
publisher = {Association for Computing Machinery},
address = {New York, NY, USA},
volume = {7},
number = {4},
issn = {0004-5411},
doi = {10.1145/321043.321046},
journal = {J. ACM},
month = oct,
pages = {326–329},
numpages = {4}
}

@article{liu2022review,
  title={A Review of Collaborative Air-Ground Robots Research},
  author={Liu, C. and Zhao, J. and Sun, N.},
  journal={Journal of Intelligent \& Robotic Systems},
  volume={106},
  number={60},
  pages={1--18},
  year={2022},
  publisher={Springer},
  doi={10.1007/s10846-022-01756-4}
}

@article{doi:10.1177/1729881417750787,
author = {Tomas Lazna and Petr Gabrlik and Tomas Jilek and Ludek Zalud},
title ={Cooperation between an unmanned aerial vehicle and an unmanned ground vehicle in highly accurate localization of gamma radiation hotspots},
journal = {International Journal of Advanced Robotic Systems},
volume = {15},
number = {1},
pages = {1729881417750787},
year = {2018},
doi = {10.1177/1729881417750787},
}

@ARTICLE{9448603,
  author={De Petrillo, Matteo and Beard, Jared and Gu, Yu and Gross, Jason N.},
  journal={IEEE Aerospace and Electronic Systems Magazine}, 
  title={Search Planning of a UAV/UGV Team With Localization Uncertainty in a Subterranean Environment}, 
  year={2021},
  volume={36},
  number={6},
  pages={6-16},
  doi={10.1109/MAES.2021.3065041}}

@InProceedings{10.1007/978-981-16-9492-9_334,
author="Ren, Shuangyin
and Chen, Rongbing
and Gao, Wei",
editor="Wu, Meiping
and Niu, Yifeng
and Gu, Mancang
and Cheng, Jin",
title="A UAV UGV Collaboration Paradigm Based on Situation Awareness: Framework and Simulation",
booktitle="Proceedings of 2021 International Conference on Autonomous Unmanned Systems (ICAUS 2021)",
year="2022",
publisher="Springer Singapore",
address="Singapore",
pages="3398--3406",
isbn="978-981-16-9492-9",
doi="https://doi.org/10.1007/978-981-16-9492-9_334"
}

@INPROCEEDINGS{7320804,
  author={Klodt, Lukas and Khodaverdian, Saman and Willert, Volker},
  booktitle={2015 IEEE Conference on Control Applications (CCA)}, 
  title={Motion control for UAV-UGV cooperation with visibility constraint}, 
  year={2015},
  volume={},
  number={},
  pages={1379-1385},
  doi={10.1109/CCA.2015.7320804}}

@article{liang2022survey,
  author    = {Yujie Liang and Zhaoxia Luo},
  title     = {A Survey of Truck--Drone Routing Problem: Literature Review and Research Prospects},
  journal   = {Journal of the Operations Research Society of China},
  volume    = {10},
  number    = {2},
  pages     = {343--377},
  year      = {2022},
  doi       = {10.1007/s40305-021-00383-4},
  publisher = {Springer},
}

@article{DEFREITAS201895,
title = {A Randomized Variable Neighborhood Descent Heuristic to Solve the Flying Sidekick Traveling Salesman Problem},
journal = {Electronic Notes in Discrete Mathematics},
volume = {66},
pages = {95-102},
year = {2018},
note = {5th International Conference on Variable Neighborhood Search},
issn = {1571-0653},
doi = {https://doi.org/10.1016/j.endm.2018.03.013},
author = {Júlia Cária {de Freitas} and Puca Huachi Vaz Penna},
}

@article{CAO2020116925,
title = {Development and uncertainty analysis of radionuclide atmospheric dispersion modeling codes based on Gaussian plume model},
journal = {Energy},
volume = {194},
pages = {116925},
year = {2020},
issn = {0360-5442},
doi = {https://doi.org/10.1016/j.energy.2020.116925},
author = {Bo Cao and Weijie Cui and Chao Chen and Yixue Chen},
}

@article{EMST-2022-0009,
title = {Atmospheric dispersion prediction of accidental release: A review},
journal = {Emergency Management Science and Technology},
volume = {2},
number = {EMST-2022-0009},
pages = {1},
year = {2022},
issn = {2832-448X},
doi = {10.48130/EMST-2022-0009},
author = { Zhan Dou and Zhe Liu and Lili Li and Hang Zhou and Qianlin Wang and Jianwen Zhang and Liangchao Chen},
}
\end{document}